\documentclass[a4paper,fleqn,usenatbib]{mnras}
\hypersetup{final=true}

\usepackage{newtxtext,newtxmath}

\usepackage[T1]{fontenc}
\usepackage{ae,aecompl}

\usepackage[utf8]{inputenc}

\usepackage{graphicx}	
\usepackage{amsmath}	
\usepackage{amssymb}	
\usepackage{etoolbox}

\usepackage{color}

\title[Spiral arms in HD100453]{Spiral arms in the proto-planetary disc HD100453 detected with ALMA: evidence for binary-disc interaction and a vertical temperature gradient}

\author[G. P. Rosotti et al]{G. P. Rosotti$^{1,2}$\thanks{rosotti@strw.leidenuniv.nl}, M. Benisty$^{3,4}$, A. Juh\'asz$^{2}$, R. Teague$^{5,6}$, C. Clarke$^{2}$, C. Dominik$^{7}$, \newauthor C. P. Dullemond$^{8}$, P. D. Klaassen$^{9}$, L. Matrà$^{6}$, T. Stolker$^{10}$\\
$^{1}$Leiden Observatory, Leiden University, P.O.~Box 9513, NL-2300~RA Leiden, the Netherlands\\
$^{2}$Institute of Astronomy, University of Cambridge, Madingley Road, Cambridge CB3 OHA, UK\\
$^{3}$Unidad Mixta Internacional Franco-Chilena de Astronom\'{i}a (CNRS, UMI 3386), Departamento de Astronom\'{i}a, Universidad de
Chile,\\ Camino El Observatorio 1515, Las Condes, Santiago, Chile\\
$^{4}$Univ. Grenoble Alpes, CNRS, IPAG, 38000 Grenoble, France\\
$^{5}$Department of Astronomy, University of Michigan, 1085 South University Avenue, Ann Arbor, MI 48109, USA\\
$^{6}$Harvard-Smithsonian Center for Astrophysics, 60 Garden Street, Cambridge, MA 02138, USA\\
$^{7}$Anton Pannekoek Institute for Astronomy, University of Amsterdam, Science Park 904,1098XH Amsterdam, The Netherlands\\
$^{8}$Institut für Theoretische Astrophysik, Universität Heidelberg, Albert-Ueberle-Str. 2, 69120 Heidelberg, Germany\\
$^{9}$UK Astronomy Technology Centre, Royal Observatory Edinburgh, Blackford Hill, Edinburgh, EH9 3HJ, UK\\
$^{10}$Institute for Particle Physics and Astrophysics, ETH Zurich, Wolfgang-Pauli-Strasse 27, 8093 Zurich, Switzerland
}

\date{Accepted 2019 October 31. Received 2019 October 31; in original form 2019 June 28}

\pubyear{2019}

\begin{document}
\label{firstpage}
\pagerange{\pageref{firstpage}--\pageref{lastpage}}
\maketitle

\begin{abstract}
Scattered light high-resolution imaging of the proto-planetary disc orbiting HD100453 shows two symmetric spiral arms, possibly launched by an external stellar companion. In this paper we present new, sensitive high-resolution ($\sim$30\,mas) Band 7 ALMA observations of this source. This is the first source where we find counterparts in the sub-mm continuum to both scattered light spirals. The CO J=3-2 emission line also shows two spiral arms; in this case they can be traced over a more extended radial range, indicating that the southern spiral arm connects to the companion position. This is clear evidence that the companion is responsible for launching the spirals. 
The pitch angle of the sub-millimeter continuum spirals ($\sim$6\degr{}) is lower than the one in scattered light ($\sim$16\degr{}). We show that hydrodynamical simulations of binary-disc interaction can account for the difference in pitch angle only if one takes into account that the midplane is colder than the upper layers of the disc, as expected for the case of externally irradiated discs.
\end{abstract}




\section{Introduction}
\label{sec:introduction}

Thanks to new, high-resolution instruments  (e.g. SPHERE/VLT, GPI/Gemini, ALMA), we can now study proto-planetary discs at unprecedented detail. The recent findings of these telescopes show that, when discs are imaged at high spatial resolutions of a few astronomical units (au), they all show conspicuous sub-structure, such as rings \citep{alma-partnership_2015,2018ApJ...869...17L,2018ApJ...869L..42H,2018ApJ...866L...6C}, crescents \citep{Casassus2013,van_der_marel_2013} and spirals \citep[e.g.,][]{garufi_2013,Christiaens2014,benisty_2015,benisty_2017,stolker_2016,perez_2016,huang2018b}. In this paper we focus on the latter. The well-studied observed spirals (e.g., MWC758, HD135344B, HD100453) have similar morphologies: two symmetric arms, shifted in azimuth by approximately 180\degr, at distances of tens of au from the star. In most cases, there is also a gap/cavity inwards of the spirals.

It is tempting to interpret the observed spiral arms as due to the presence of young planets lurking in these discs: through their gravitational influence, planets perturb their natal discs \citep{goldreich_1979} and excite spiral arms in discs. If the planet interpretation is correct, the spirals arms could be used as planet signposts, allowing us to study the young exoplanetary population. However, often the spiral morphology as predicted by models corresponds to a single spiral arm \citep{ogilvie_2002}, in contrast to most observations that show two symmetric arms. This led \citet{juhasz_2015} to argue, using numerical simulations of planets orbiting \textit{inwards} of the spirals, that the observed spirals cannot be produced by planets. More recently, \citet{dong_2015a} and \citet{Zhu2015} highlighted that the spiral morphology is in fact compatible with the presence of planets, provided that they are massive and orbiting \textit{outwards} of the spirals (i.e., at greater distances from the star), as under these two conditions the number of spiral arms increases to two (see \citealt{2018ApJ...859..119B} for a recent, more comprehensive survey of the parameter space).

However, the planet interpretation is not unique. A natural, alternative explanation is the self-gravity of the disc \citep[e.g.,][]{rice_2003,Cossins2010}; although this typically produces many spirals arms in the disc \citep{cossins_2009}, with the exact number depending on the Toomre Q parameter \citep{toomre_1964}, there are regions of the parameter space (namely for massive discs) where only two are produced \citep{Hall2016}. Self-gravity is for example a good explanation for the spirals observed in Elias 2-27 \citep{perez_2016}, as shown by \citet{meru_2017}; see also \citet{2018MNRAS.477.1004H,2018ApJ...860L...5F}. It is also possible that the finite telescope resolution detects only two spirals, even if more are present \citep{2014MNRAS.444.1919D}. Moreover, even if the disc is only close to the self-gravitating regime, the presence of a planet might tip the balance and trigger self-gravitating spiral arms \citep{pohl_2015}. While most discs would require uncomfortably high masses to explain the spiral morphology by self-gravity, this is nevertheless a possibility that cannot be excluded. There are also other alternative explanations for spiral arms, such as for example the finite light travel time from the star \citep{2016A&A...593L..20K} or the presence of shadows \citep{Montesinos2016}.

Initially, all the known spiral arms had been observed only in scattered light. While an intriguing finding,  
scattered light observations only trace the surface layers, preventing us from confirming whether the spiral arms extend all the way to the midplane or if they are features only in the atmosphere of the disc. More recently (after the initial findings, e.g. \citealt{Tang2012,Christiaens2014}, of spiral structures in molecular tracers coming from the upper layers), ALMA has observed spirals in proto-planetary discs in the continuum emission originating in the midplane, for example around the T Tauri stars Elias 2-27 \citep{perez_2016}, IM Lup, and WaOph 6 \citep{huang2018b}, around the massive star G17.64+0.16 \citep{Maud2019} and around the intermediate mass star MWC 758 \citep{Boehler2018,Dong2018}. The latter source is known to show prominent spiral arms in scattered light, in principle allowing a multi-wavelength comparison of the spiral morphology; however the comparison is not straightforward because the sub-mm image only shows one spiral arm, while the scattered light image shows two.

\citet{juhasz_2018} have shown that a multi-wavelength comparison is particularly interesting because the two wavelengths trace different layers of the disc: the sub-millimeter continuum emission originates from the midplane while the scattered light probes the disc upper layers. The upper layers are generally hotter\footnote{These constraints come from dust radiative transfer, while the spiral pitch angles depend on the gas temperature. A confirmation of different pitch angles in the two tracers is therefore also a confirmation of good thermal coupling between gas and dust.} in passively heated discs \citep[e.g.,][]{Calvet1991}, which leads (as expected theoretically and shown by hydrodynamic simulations) to higher pitch angles of the planetary spirals in scattered light compared to the sub-mm (see \citealt{Lee2015} for a general study of wave propagation in a thermally stratified disc). As well as a probe of the disc vertical thermal structure, this difference is also a test of the planetary hypothesis. While a quantitative study similar to \citet{juhasz_2018} has not been performed for spirals produced by gravitational instability, in this case the disc midplane is heated by shocks that tend to erase  or invert the temperature difference with the upper layers \citep[e.g.,][]{2002ApJ...567L.149B}. Therefore, in this case no significant difference in pitch angle between observational tracers is expected.

In this context, HD100453 represents a unique system to study. Scattered light imaging has shown  that the system presents 2 symmetric spiral arms \citep{wagner_2015,benisty_2017}. However, in contrast to all other discs with spirals, in this case the central star has a known stellar companion. HD100453A has a spectral type A9V \citep{Dominik2003} and a mass of 1.5 $M_\odot \pm 0.15$ \citep{Fairlamb}, while the companion is an M dwarf companion with a mass of $\sim$ 0.2 $M_\odot$. The projected separation is 1.05\arcsec{} \citep{Chen2006,Collins2009}. The companion has been confirmed to be co-moving and more recently the orbital parameters of the binary have been constrained \citep{2018ApJ...854..130W} using direct imaging observations at different epochs. Hydrodynamical simulations support the hypothesis that the companion launches spirals with properties compatible with those observed \citep{dong_2016}. Recently this picture has been challenged by the ALMA Band 6 observations of \citet{vanderplas2019}, which show an extended disc in CO emission around the primary, extending almost up to the companion location, seemingly in contradiction with the standard picture of disc truncation by companions \citep{artymowicz1994}. In addition, the data did not show a clear sub-mm counterpart to the scattered light spirals. Motivated by these findings, the authors proposed that the companion might be on an inclined orbit and not responsible for launching the spirals, arguing instead for the shadow origin \citep{Montesinos2016}.

Given these controversies, HD100453 is a unique laboratory to test models of spiral launching mechanisms and compare models of binary-disc interaction with observations. In this paper, we present new high-resolution ($\sim$30\,milli-arcseconds) ALMA observations of the source, showing that the scattered light spirals have in fact clear sub-mm counterparts both in the continuum and gas CO emission. We thus confirm that the spirals are actual structures in the disc surface density and not only in the surface layers of the disc; because one of the two spiral arms detected in CO emission points to the location of the companion, we also confirm that the companion is responsible for launching the spirals. We then use these observations to test the hypothesis of \citet{juhasz_2018} that the spiral pitch angle should depend on the tracer.

The paper is structured as follows: we first present the observational results in Section \ref{sec:observation}. We then present the results of a simple geometrical model in Section \ref{sec:modelling} to prove that the different observed spiral pitch angle between ALMA and SPHERE do not result from a projection effect. We show in Section \ref{sec:hydro_data_comparison} that hydrodynamical simulations of planet-disc interaction can account for the observed difference in pitch angles. We discuss the limitations of our models, possible alternative scenarios and the importance of our results in the context of planet-disc interaction and the sub-structure observed in other discs in Section \ref{sec:discussion}. We finally draw our conclusions in Section \ref{sec:conclusions}.

\section{Observations}
\label{sec:observation}
HD100453 was observed with the Atacama Large Millimetre/submillimetre Array (ALMA) on the 24th and 25th November 2017 (Project ID 2017.1.01424.S, PI: A.~Juh\'asz). Our target was observed with 40 antennas with baselines ranging from 92\,m to 8547\,m, and the
total on-source integration time was 1h 46min. Formally, the maximum recoverable scale with this antenna configuration is 0.6\arcsec{}. Given that this is slightly smaller than the companion separation, in principle we might be missing information on the largest spatial scales. We will discuss this concern in the following sections. The correlator was set up to use four spectral windows in Band 7, centred on 
345.79599\,GHz, 343.810092\,GHz, 331.80974\,GHz and 333.809798\,GHz, respectively. The first spectral window, centred
on 345.79599\,GHz, was set to Frequency Division Mode (FDM) with a channel spacing of 488.281\,kHz, corresponding to
0.84\,km/s velocity resolution after Hanning smoothing, to observe the CO J=3-2 line. The remaining three
spectral windows were set to Time Division Mode (TDM) to observe the continuum. All four spectral windows had a bandwidth
of 1.875\,GHz. To calibrate the visibilities we used the ALMA pipeline and the Common Astronomy Software Applications (CASA, version 5.1.1; \citealt{2007ASPC..376..127M}). Since self-calibration of both continuum and gas data did not significantly improve the images, we will base our analysis on the non-self-calibrated data.

The calibrated visibilities were imaged using the {\tt clean} task in CASA. For what concerns the continuum, we used Briggs weighting with a robust parameter value of 0.5, achieving a resolution of 0.036\arcsec{} x 0.031\arcsec{} and a beam position angle of -38.36\degr. The rms noise level in the continuum was 
22\,$\mu$Jy/beam. While imaging the CO emission at this resolution still recovers the emission from the disc, a clear detection of the southern spiral up to the companion position (see later) requires to sacrifice some spatial resolution in exchange for surface brightness sensitivity. Therefore, in all the plots shown in this paper for the CO emission we used a robust parameter of 2 (corresponding to natural weighting); in this case the spatial resolution was 0.054\arcsec{} x 0.052\arcsec{} with a beam position angle of 83\degr{}. The rms noise level was 0.95\,mJy/beam in a single channel.

\section{Results}

\subsection{Continuum}
\label{subsec:cont_image}

\begin{figure*}
\center
\includegraphics[width=\columnwidth]{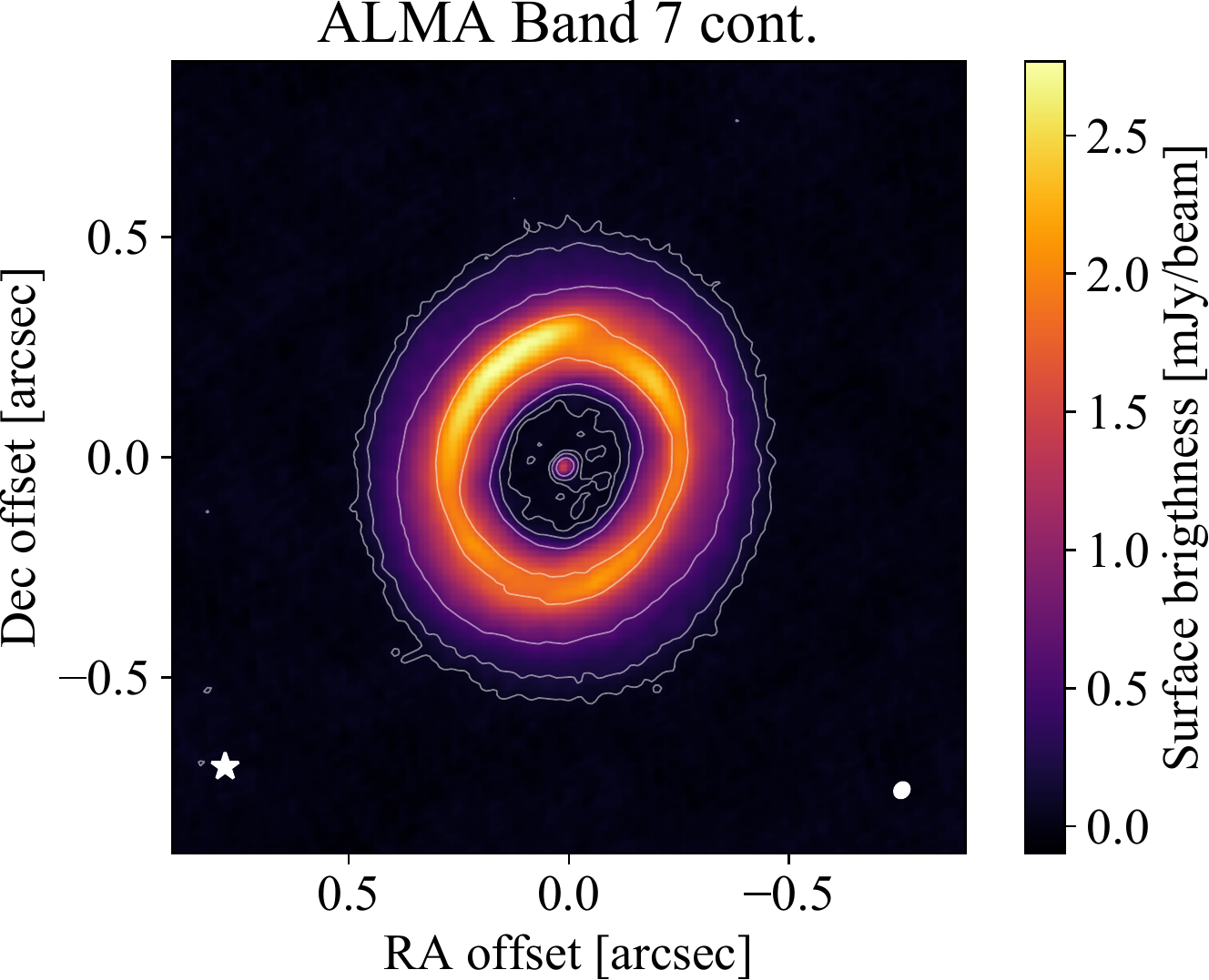}
\includegraphics[width=\columnwidth]{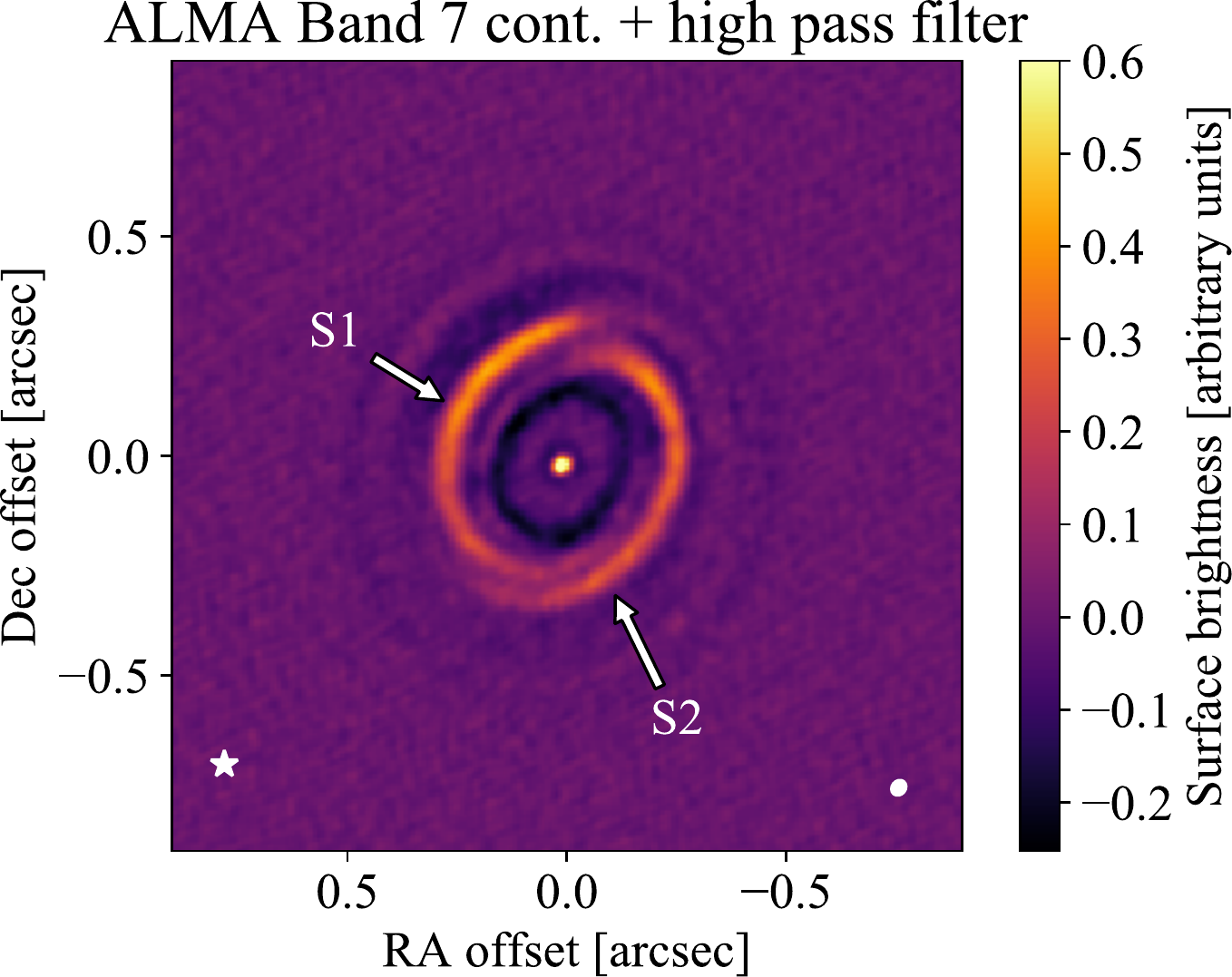}
\caption{{\it Left:} ALMA Band 7 continuum image. The white ellipse in the bottom right corner shows the synthesised beam while the white star marks the projected position of HD100453B. The contours show 3$\sigma$, 9$\sigma$, 27$\sigma$ and 81$\sigma$ levels. {\it Right:} ALMA Band 7 continuum image after a high pass filter has been applied in the image plane. Two spiral arms, marked S1 and S2 in the figure, are clearly visible in the filtered image.}
\label{fig:continuum_image}
\end{figure*}

The observed continuum image is presented in the left panel of \autoref{fig:continuum_image}. We measure a total continuum flux in a circular aperture encompassing the extent of the ring of 510\,mJy. The image shows a ring of emission
between 0.15\arcsec{}~and 0.51\arcsec{}, an inner cavity inwards of about 0.15\arcsec{} and emission at the centre of the disc. This emission is likely unresolved (a Gaussian fitting gives an unconvolved size smaller than 1/3 of the beam) and has a flux of $\sim 1.6$\,mJy. 
While the surface brightness distribution along the ring is not azimuthally symmetric, there is no obvious counterpart to the shadows observed in scattered light. The surface brightness peaks at a position
angle of about 40\degr, while it shows a dip at position angles of around -10\degr~and 170\degr.

To further investigate the nature of the asymmetry we applied a high pass filter to the image, convolving it with an appropriate filter kernel. We chose an inverse Gaussian filter kernel, which we defined in the Fourier space as
\begin{equation}
K(\nu) = 1.0 - \exp{\left(-\frac{\nu^2}{2\sigma_\nu^2}\right)}
\end{equation}
where $\nu$ is the spatial frequency and $\sigma$ is the width of the filter which we took to be 0.2\,arcsec$^{-1}$. The filtered image is shown in the right panel of \autoref{fig:continuum_image}. With the large scale emission removed, the
high pass filtered image reveals two spiral arms. The S1 arm extends from PA=$\sim$0\degr~to PA=$\sim$200\degr, while the 
S2 arm extends from PA=$\sim$160\degr, PA=$\sim$360\degr.

\subsection{CO J=3-2}
\label{subsec:co_image}

\begin{figure*}
\center
\includegraphics[width=\columnwidth]{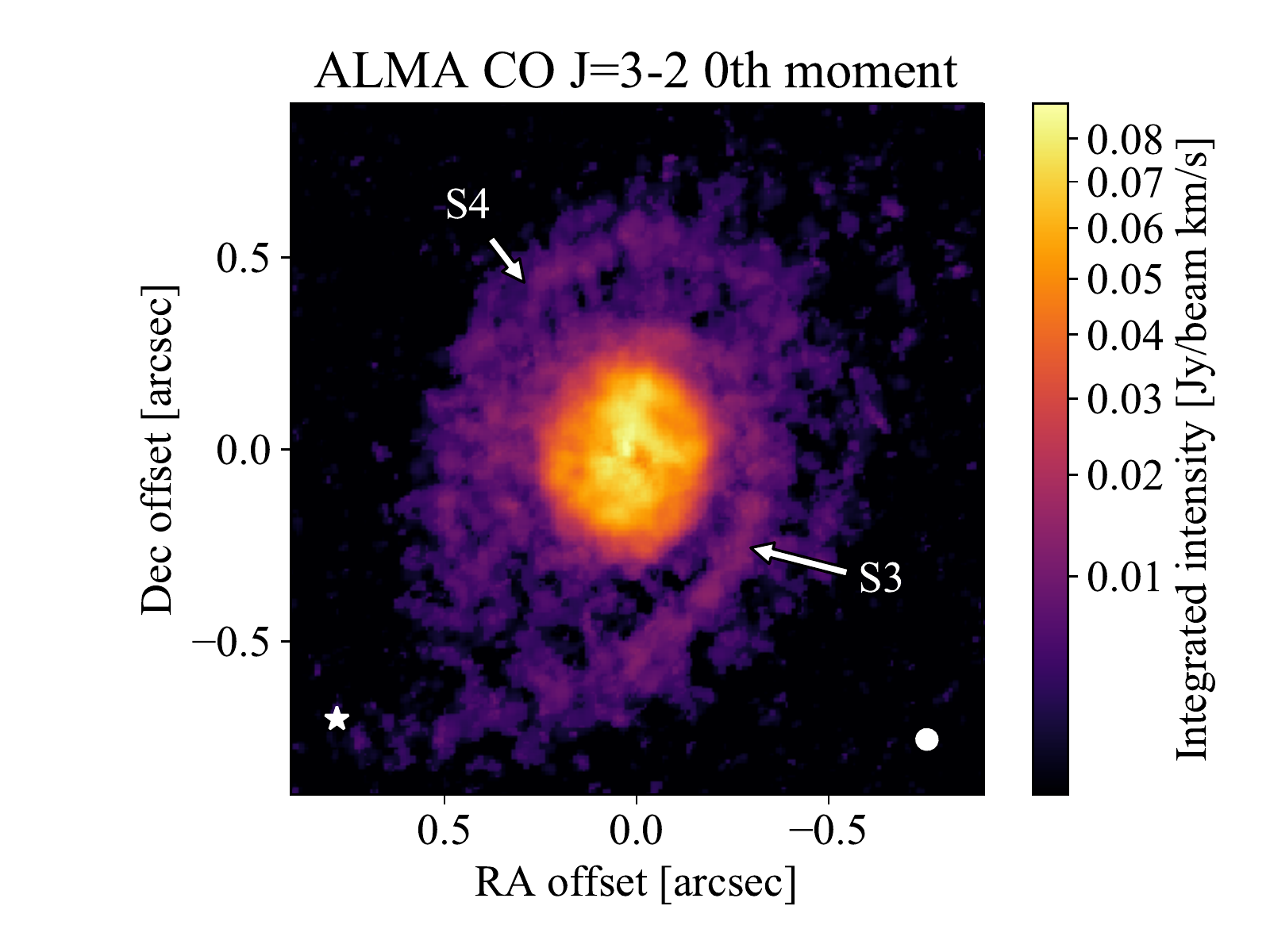}
\includegraphics[width=\columnwidth]{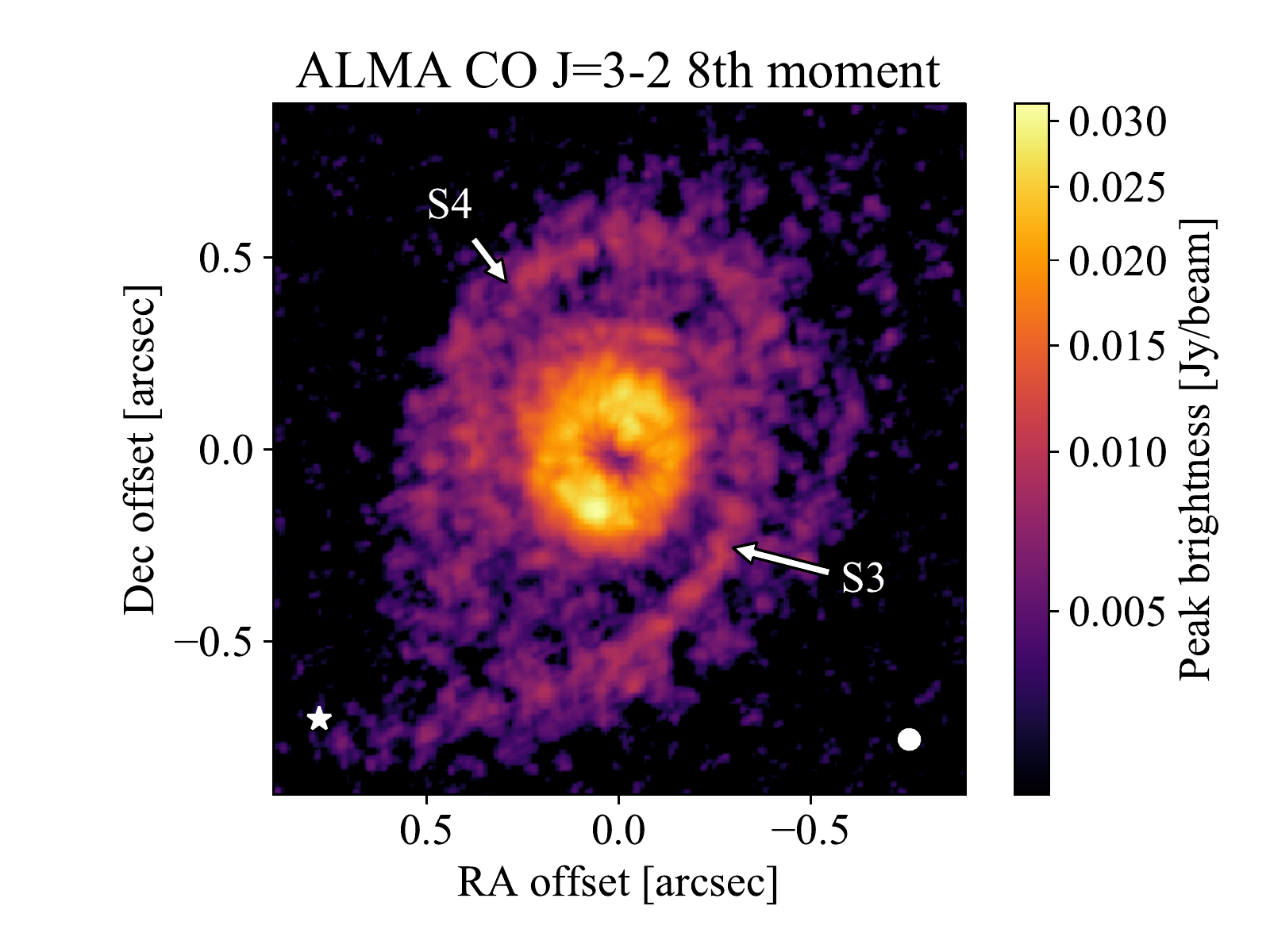}
\caption{{\it Left:} CO J=3-2 integrated intensity (0th moment) map.  {\it Right:} CO J=3-2 peak intensity (8th moment) map. 
No gap or hole is detected in the gas, which extends inwards all the way to the innermost resolution element. Like in the continuum, two spirals, marked S3 and S4 in the figure, are very clearly visible in both the integrated intensity and the peak intensity map, although at larger distances from the star (see also later \autoref{fig:cont_co_overaly}). The southern spiral S3 points to the location of the companion.}
\label{fig:co_integrated_intensity}
\end{figure*}

The left panel of \autoref{fig:co_integrated_intensity} presents the CO J=3-2 integrated intensity map while the right panel of \autoref{fig:co_integrated_intensity}  presents the peak intensity (8th moment) map. Both maps show disc-like emission out to about $0.33\arcsec{}$ from the centre of the disc, without a visible hole or central depression; while there is a lack of emission at the centre of the disc in the peak brightness map, this is merely a consequence of the Keplerian shear and finite spatial resolution (see e.g. discussion in \citealt{Huang2018TwHya}). Interestingly, the images also reveal two large scale spirals, S3 and S4, extending 
from the outer edge of the disc at $0.33\arcsec{}$  to about $0.6\arcsec{}$  in radial distance from the centre of the disc. The southern spiral points to the position of the companion, lending support to the hypothesis that the spirals are launched by the companion \citep{dong_2016}.

\begin{figure}
\center
\includegraphics[width=\columnwidth]{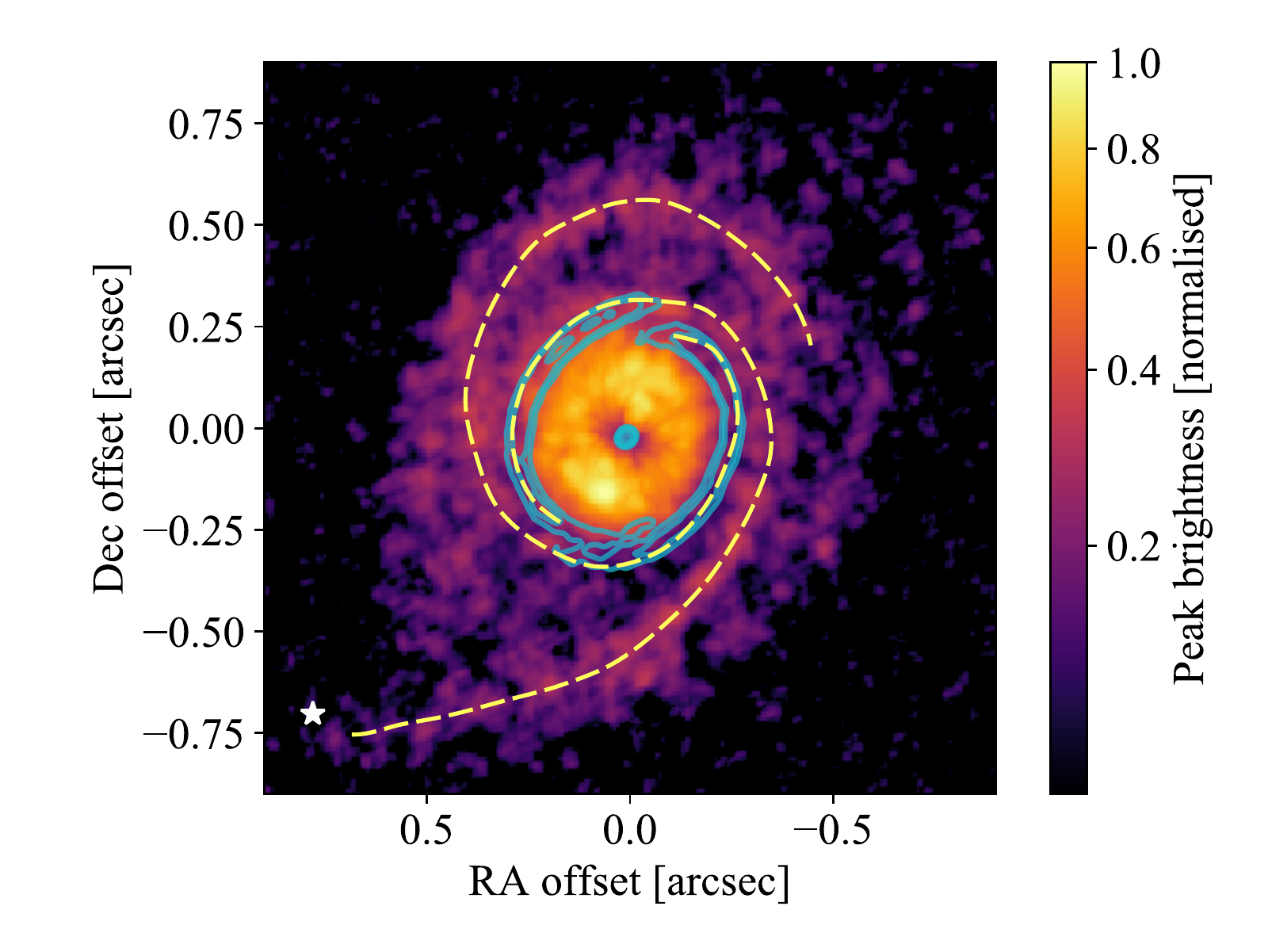}
\caption{Overlay of the ALMA B7 continuum image (blue contours) on the CO J=3-2 peak intensity (8th moment) map. The spirals in the continuum and in the CO
seems to be well aligned (continuum S1 to CO S3 and continuum S2 to CO S4) as if they would be two parts of the same spirals. The yellow dashed lines are a visual guide to highlight this connection.}
\label{fig:cont_co_overaly}
\end{figure}

To better highlight the connection between the spirals in CO and the spirals in the continuum, in \autoref{fig:cont_co_overaly} we overlay the continuum image on top of the CO peak intensity map. The spirals 
in CO start approximately at the position where the spiral arms in the continuum end.  The alignment of the spirals in CO and continuum tentatively suggests that S1 and S3 are two parts of the same spiral density wave, and so are S2 and S4.
\begin{figure*}
\center
\includegraphics[width=0.8\textwidth]{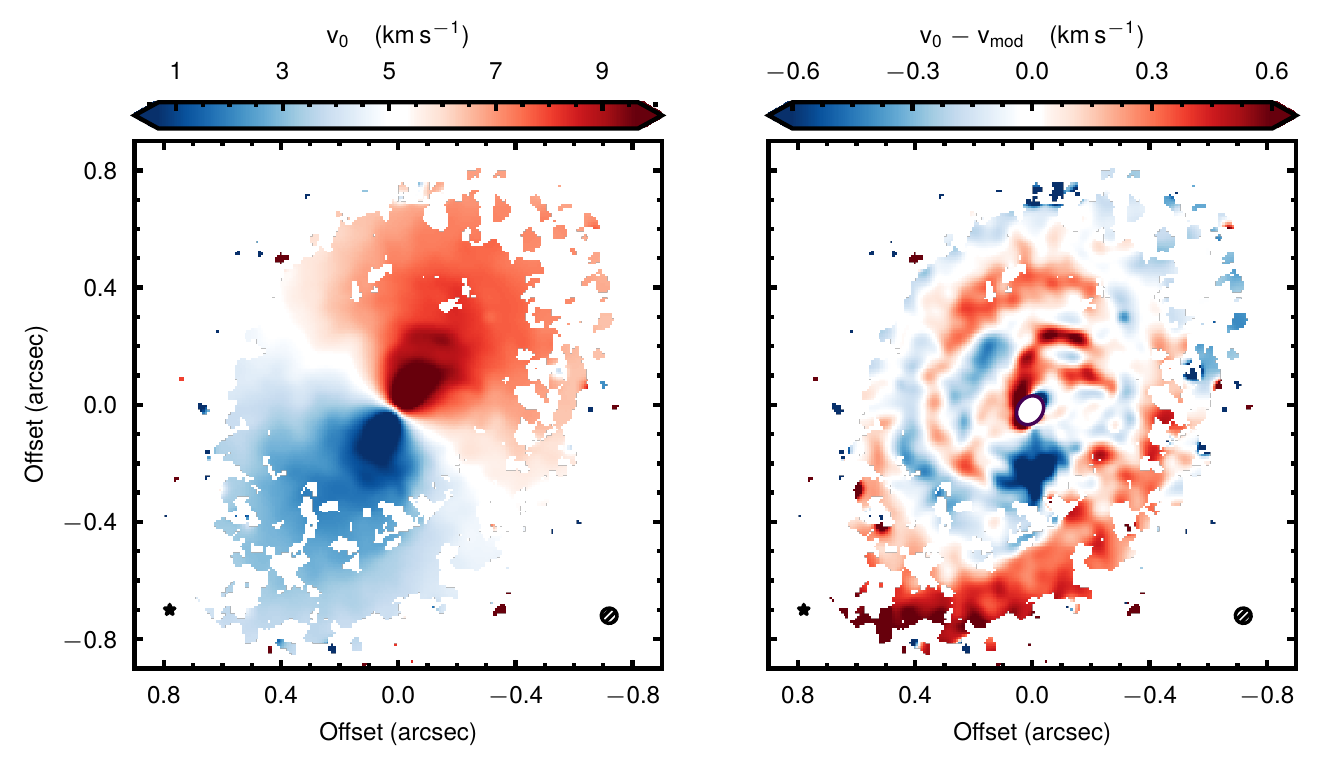}
\caption{\textit{Left}: CO J=3-2 projected velocity (1st moment) map. \textit{Right}: residuals of the fit to the projected velocity map.}
\label{fig:projected_velocity}
\end{figure*}

In \autoref{fig:projected_velocity} we present the projected velocity (1st moment) map\footnote{Computed using \texttt{bettermoments} \citep{bettermoments}}. The map shows Keplerian rotation with the North-Western part of the disc moving away from the observer, and the South-Eastern part of the disc rotating towards the observer.  Qualitatively, the Keplerian velocity pattern is retained along the spirals S3 and S4 confirming that these spiral arms are indeed part of the disc and bound to HD100453. As already noted by \citet{2018ApJ...854..130W} and \citet{vanderplas2019}, since the south side of the disc is blueshifted and the morphology of the scattered light emission shows that it is the near side (see discussion in \citealt{benisty_2017}), the disc rotates counter-clockwise in the plane of the sky. Therefore, the spirals are trailing and compatible with the companion origin.

\subsection{Comparison with scattered light observations}
\label{sec:comparison_alma_sphere}

\begin{figure}
\center
\includegraphics[width=8.6cm]{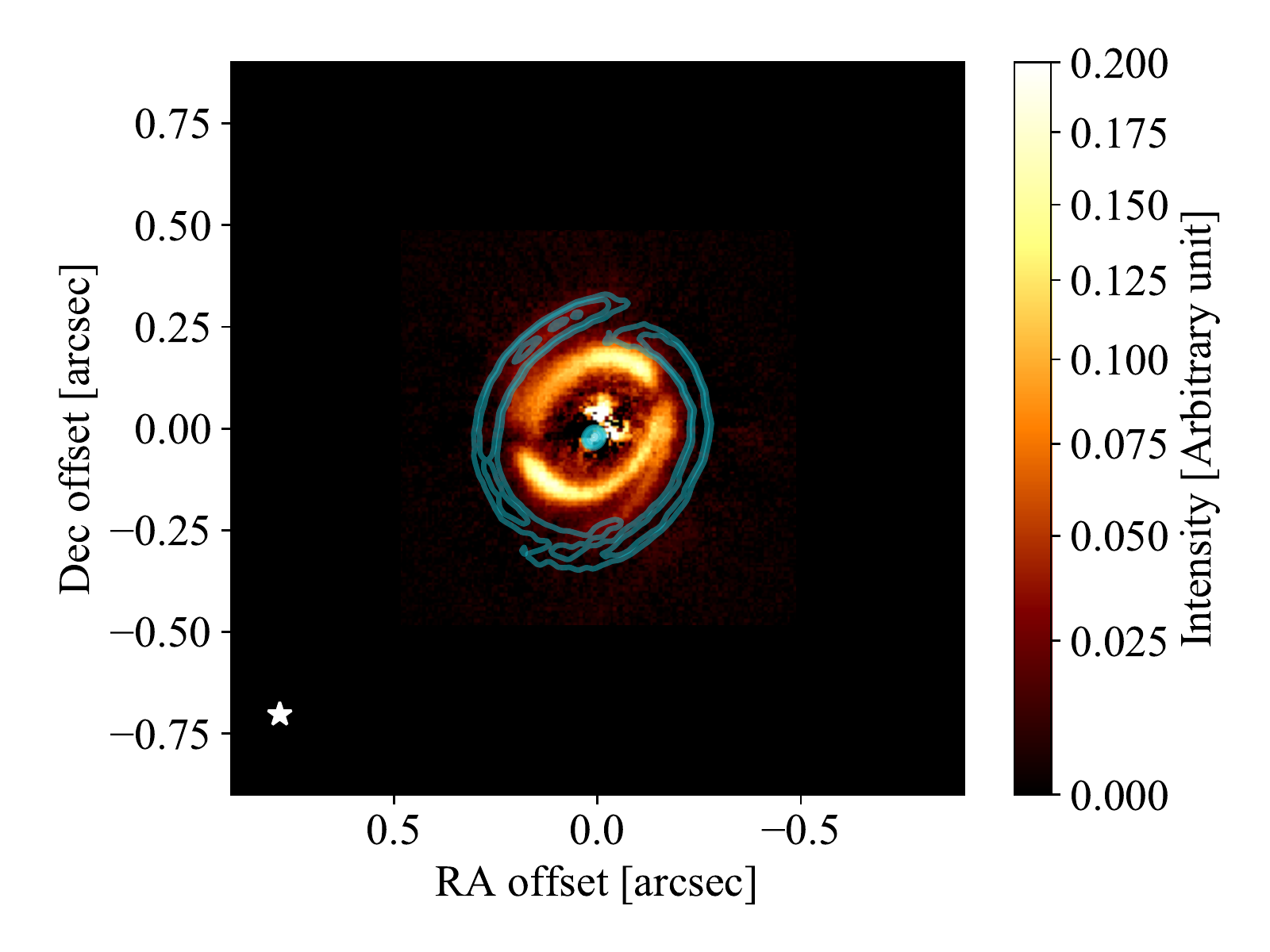}
\caption{Overlay of the ALMA B7 continuum image (blue contours) on the SPHERE R' image from \citet{benisty_2017}. The spiral arms in scattered light emission have a larger pitch angle than those in the sub-mm continuum.}
\label{fig:sphere_alma_cont_overlay}
\end{figure}

\citet{juhasz_2018} suggested that the pitch angle of spiral density waves in proto-planetary discs with vertical thermal stratification depends on the
vertical temperature profile. Therefore the pitch angle of spirals in passively irradiated discs, with positive vertical temperature
gradient, will be the lowest in the disc midplane and the highest in the disc atmosphere. To study this effect in HD100453 we present in \autoref{fig:sphere_alma_cont_overlay} a comparison between the 
ALMA Band 7 continuum image, probing the disc midplane, and the SPHERE R$^\prime$ image from \citet{benisty_2017}, probing the disc atmosphere. The images have been aligned assuming that the unresolved point source in the ALMA image traces the location of the star. The emission in the SPHERE image is slightly offset to the North-East compared to the ALMA image. This is because the emission in the ALMA image is tracing a planar surface, while the scattered light is originating in a conical surface above the disc midplane. At face value, the spirals in the near-infrared have a significantly higher pitch angle compared to the spirals in the sub-millimetre continuum; we will elaborate further on this difference in the next section.

\section{Geometrical modelling}
\label{sec:modelling}

As mentioned in Section \ref{sec:comparison_alma_sphere}, there is some offset between the ALMA and SPHERE images, because the emission is coming from different heights above the midplane. It is clear that any quantitative analysis needs to take these projection effects into account. It is particularly important to address the question of whether projection effects could account for a different \textit{observed} spiral pitch angle, even if the \textit{intrinsic} pitch angle is similar. To this end, in this section we first use (Section \ref{sec:projected_velocity}) the CO velocity map to estimate the disc inclination and position angle. We then (Section \ref{sec:geometric_model}) construct a simple geometrical model of the spirals to investigate the question mentioned above and then (Section \ref{sec:de-proj}) show the result of de-projecting the images in polar coordinates using the values of the disc inclination previously constrained. 

\subsection{Disc inclination and position angle}
\label{sec:projected_velocity}

Previous studies \citep{wagner_2015,2018ApJ...854..130W,benisty_2017,2017ApJ...838...62L,vanderplas2019} found values of the disc inclination $\sim 30-40\degr$, but only \citet{2018ApJ...854..130W} and \citet{vanderplas2019} derived these values from kinematical data rather than from the emission morphology. These two studies used data from the same project (although the former used only the low resolution part of the data set) but reached slightly different conclusions, motivating the need to confirm the inclination value from our independent data set. In particular, \citet{vanderplas2019} reported the presence of a warp in the disc; they found that dividing the disc in two parts with different inclinations, with a separation radius of 38 au, provides a better fit to the data than a monolithic disc. They report a change of inclination between the two parts of the disc of 5\degr.  

To study the disc inclination, we fitted the projected velocity map following the method in \citet{2018ApJ...868..113T} using the code \texttt{eddy}\footnote{\url{https://github.com/richteague/eddy}}. We impose a Gaussian prior on the stellar mass of 1.5 $M_\odot \pm 0.15$ \citep{Fairlamb} and we assume a distance of 103pc \citep{2016A&A...595A...1G,2018A&A...616A...1G}. The best-fit value is $35 \degr$ for the disc inclination and $145 \degr$ for the position angle, which are in good agreement with previous investigations; the fit converges to a stellar mass of 1.27 $M_\odot$. If we instead fix the stellar mass to $1.5 M_\odot$, the fit converges to a lower disc inclination of 30\degr. This shows that, without a precise knowledge of the stellar mass, it is not possible to constrain the disc inclination with a precision better than a few degrees.

\autoref{fig:projected_velocity} shows the residuals of the best fit model to the projected velocity map. The spiral arms S3 and S4 are still visible in the residuals, implying that either the motion along the spirals is not entirely Keplerian, or there is additional radial or vertical motions contributing to the line of sight velocity. Since the spirals dominate the residuals, a better description of the kinematical data would require including the spirals in the model, rather than employing azimuthally symmetrical models. For this reason we do not attempt to fit the kinematics with models including a disc warp, as suggested by \citet{vanderplas2019}. We also cannot exclude the presence of a warped inner disc at distances from the star smaller than our beam; such a disc has been invoked \citep{benisty_2017,2017A&A...604L..10M} to explain the shadows seen in the scattered light image (see section \ref{sec:origin_spirals}).

The residual map shows further structures at small separations from the star, particularly in the North-West at $\sim$0.2\arcsec{} projected separation. Similar structures have been recently claimed \citep[e.g.,][]{2019arXiv190606302C} to be evidence of planets embedded in discs. The quality of the current data however does not allow us to study further this hypothesis.

\subsection{Geometric toy model for the spiral arms}
\label{sec:geometric_model}

\begin{figure}
\center
\includegraphics[width=\columnwidth]{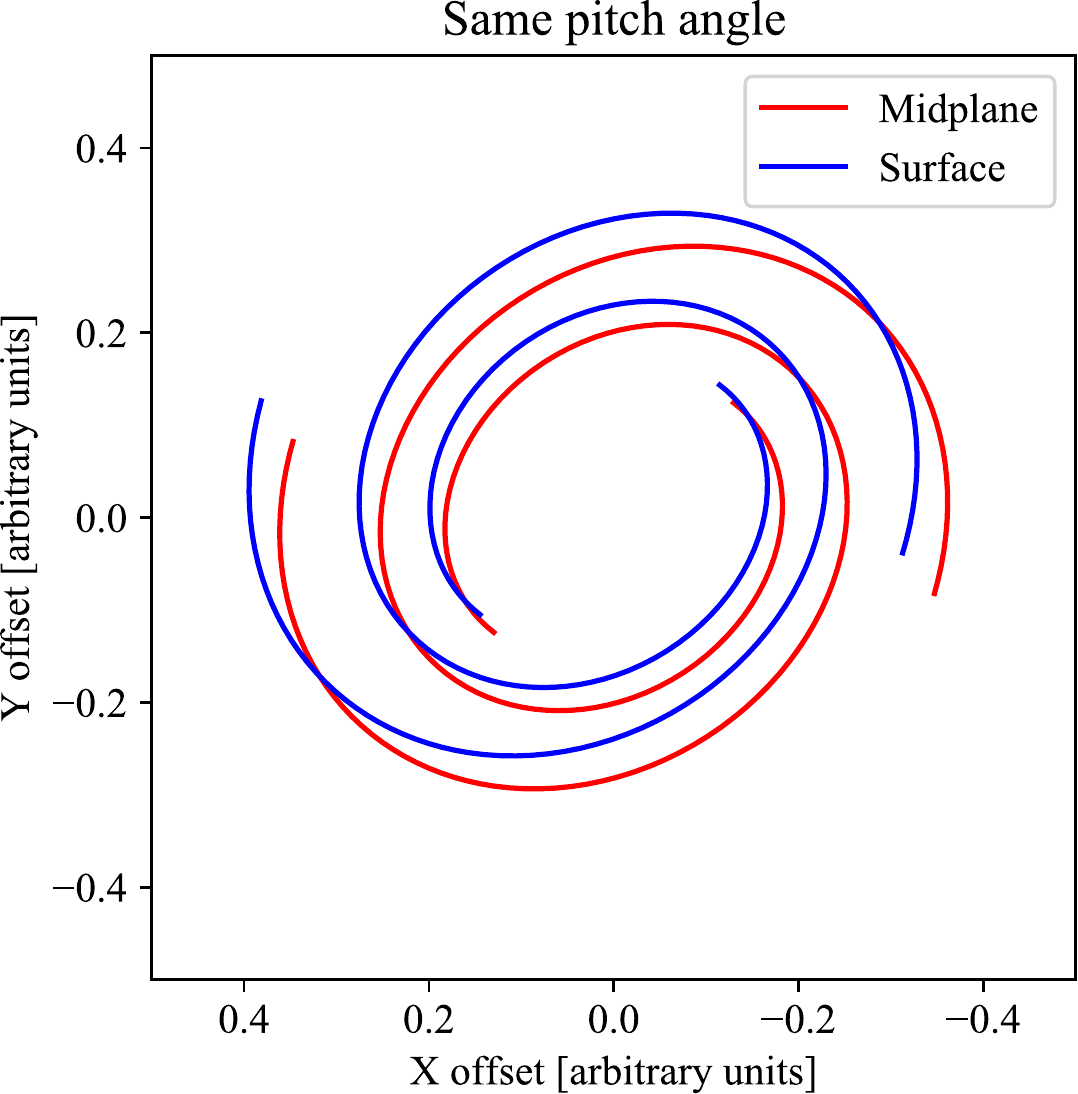}

\vspace{0.5cm}

\includegraphics[width=\columnwidth]{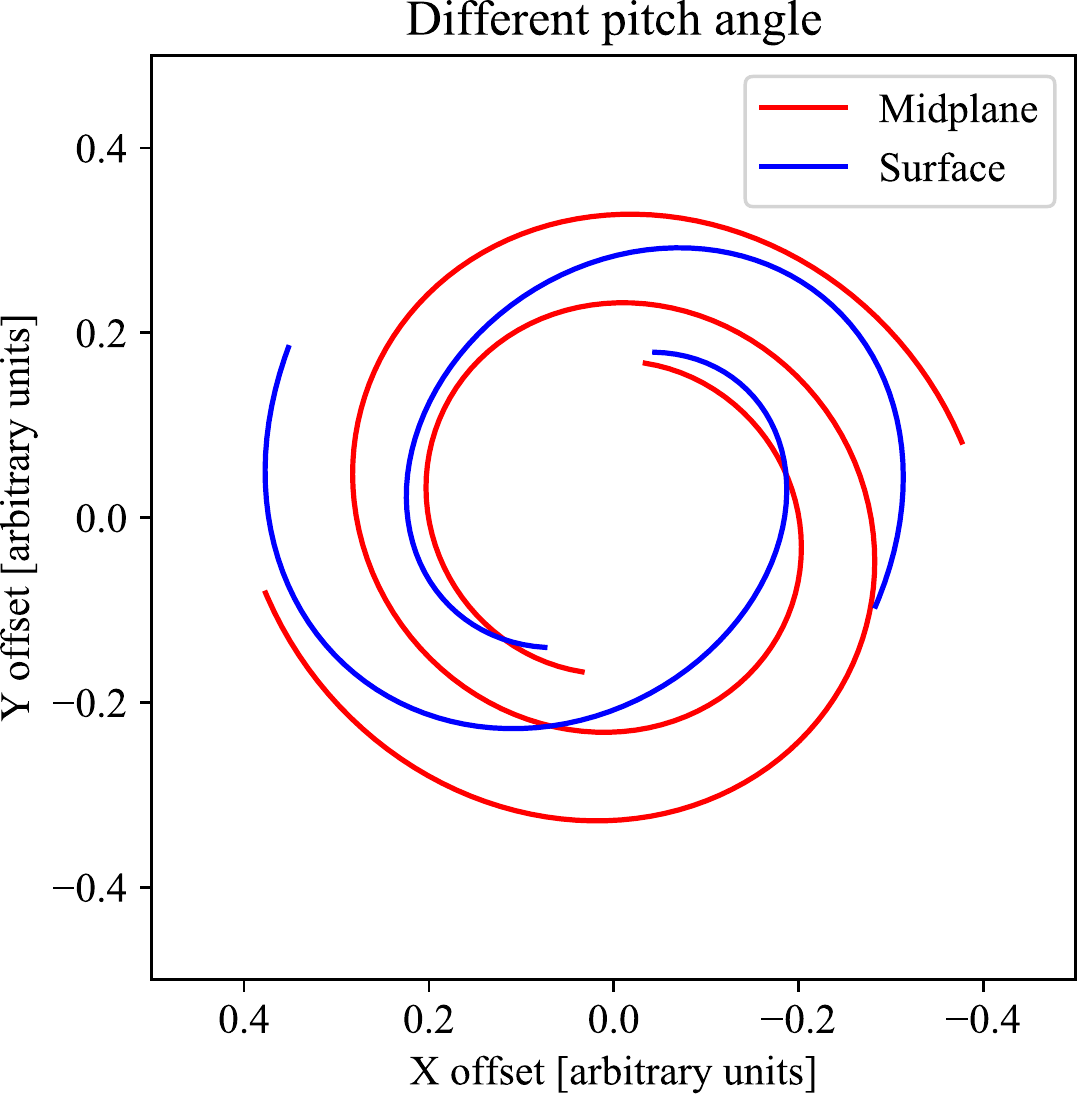}
\caption{Geometric model demonstrating the effect of projection of spiral wakes in a plane (disc midplane, red lines)  and on a conical surface (disc surface, blue lines). {\it Top:} If the disc is vertically isothermal, the spirals in the disc surface layer and in the disc midplane will only be shifted along the minor axis of the disc, but their opening angle will not change significantly (nor consistently: the change in the North-East is in the opposite direction than the change in the South-West). {\it Bottom:} If the disc has a positive vertical temperature gradient, the spirals in the surface layers will have a larger opening angle compared to the spirals in the disc midplane. }
\label{fig:cartoon}
\end{figure}

The purpose of this section is to establish whether projection effects, namely the fact that scattered light signal is coming from the upper layers of the disc (roughly a conical surface) rather than from the midplane, can cause a spurious difference in the spiral pitch angles between sub-mm and scattered light signal, as observed in the data. To investigate this possibility, we employ a simple geometric toy model, in which the 3D cartesian coordinates of the spiral wake are given by
\begin{align}
\begin{split}
x &= r\cos{[\phi(r)]}\\
y &= r\sin{[\phi(r)]}\\
z &= z_{\rm p}\left(\frac{r}{r_{\rm p}}\right)^{\rm f}
\end{split}
\end{align}
where $r$ and  $\phi(r)$ are radial and azimuthal polar coordinates in the disc midplane, $\phi(r)$ is given by the analytic wake equation for spiral waves in the linear regime by \citet{rafikov_2002a}:
 \begin{eqnarray}
\nonumber \phi(r) &&= \phi_{\rm p} - \\
 &&\nonumber \frac{{\rm sgn}(r-r_{\rm p})}{h_{\rm p}} \left(\frac{r}{r_{\rm p}} \right)^{1+\beta} \left\{\frac{1}{1+\beta} - \frac{1}{1-1.5+\beta}\left(\frac{r}{r_{\rm p}}\right)^{-1.5}\right\} \\
 &&+ \frac{{\rm sgn}(r-r_{\rm p})}{h_{\rm p}} \left( \frac{1}{1+\beta} - \frac{1}{1-1.5+\beta}\right),
\label{eq:sp_wake_muto}
\end{eqnarray}
$r_{\rm p}$ is the position of the companion and ${\rm f}$ is the flaring index. In equation \ref{eq:sp_wake_muto}, $\beta = 0.5-{\rm f}$ is the power exponent of the radial distribution of the sound speed ($c_{\rm s} \propto r^{-\beta}$), $h_{\rm p} = H_{p}(r_{\rm p})/r_{\rm p}$ is the aspect ratio of the disc at $r_{\rm p}$ and $\phi_{\rm p}$ is the azimuthal coordinate of the planet. Images of the spiral wake at any orientation are computed by applying the appropriate rotations to the spiral coordinates. The position of the companion (separation of 1.05\arcsec{}, at PA = 132\degr) was taken from \citet{wagner_2015}. The disc aspect ratio in equation \ref{eq:sp_wake_muto} should not be confused with the height of the emission surface in scattered light; in this toy model the aspect ratio in the midplane should be simply regarded as a free parameter and we use a value of 0.215. We assumed a disc inclination of 33\degr~, a position angle of 145\degr~ and a flaring index of 0.04 (these parameters are close to the ones used by \citealt{benisty_2017}  to deproject the scattered light image following the methodology of \citealt{2016A&A...596A..70S}). We artificially produce a m=2 spiral by shifting the solution of the spiral wake equation by 180\degr~in azimuth. We assumed that $z_{\rm p} = 0.22 r_{\rm p}$  to model the spirals coming from the disc surface and $z_0 = 0$ for modelling the spirals in the disc midplane.

The resulting image is presented in the top panel of \autoref{fig:cartoon}. The figure shows that there is an offset between the spiral in the midplane and the spiral coming from the upper layer, which is reminiscent of the difference seen in observations between the ALMA and the SPHERE data (see \autoref{fig:sphere_alma_cont_overlay}). The offset is maximum along the disc minor axis. In this model there is no \textit{real}  difference in pitch angle between the midplane and the surface and any \textit{apparent} difference is purely due to projection. While the differences in the apparent pitch angle are small, they do however exist. For example, in the North-East the surface spiral has a slightly higher apparent pitch angle. Note however how this reverses in the South-West, where the midplane spiral has a higher apparent pitch angle. This is different from what we see in the data, where both spirals in scattered light have a higher apparent pitch angle.

Producing a difference in pitch angle which is always in the same direction requires changing the intrinsic pitch angle of the spiral. This is shown in the bottom panel of \autoref{fig:cartoon}, where we assumed a temperature difference of a factor of 2.5 between the disc midplane and the surface layer. In this case, the surface spiral has always a higher apparent pitch angle than the midplane spiral.

We conclude that the difference in pitch angles between the SPHERE and the ALMA data cannot be explained as a projection effect and requires an intrinsic difference in pitch angle to be explained.

\subsection{De-projected images}
\label{sec:de-proj}

\begin{figure*}
\center
\includegraphics[width=16.cm]{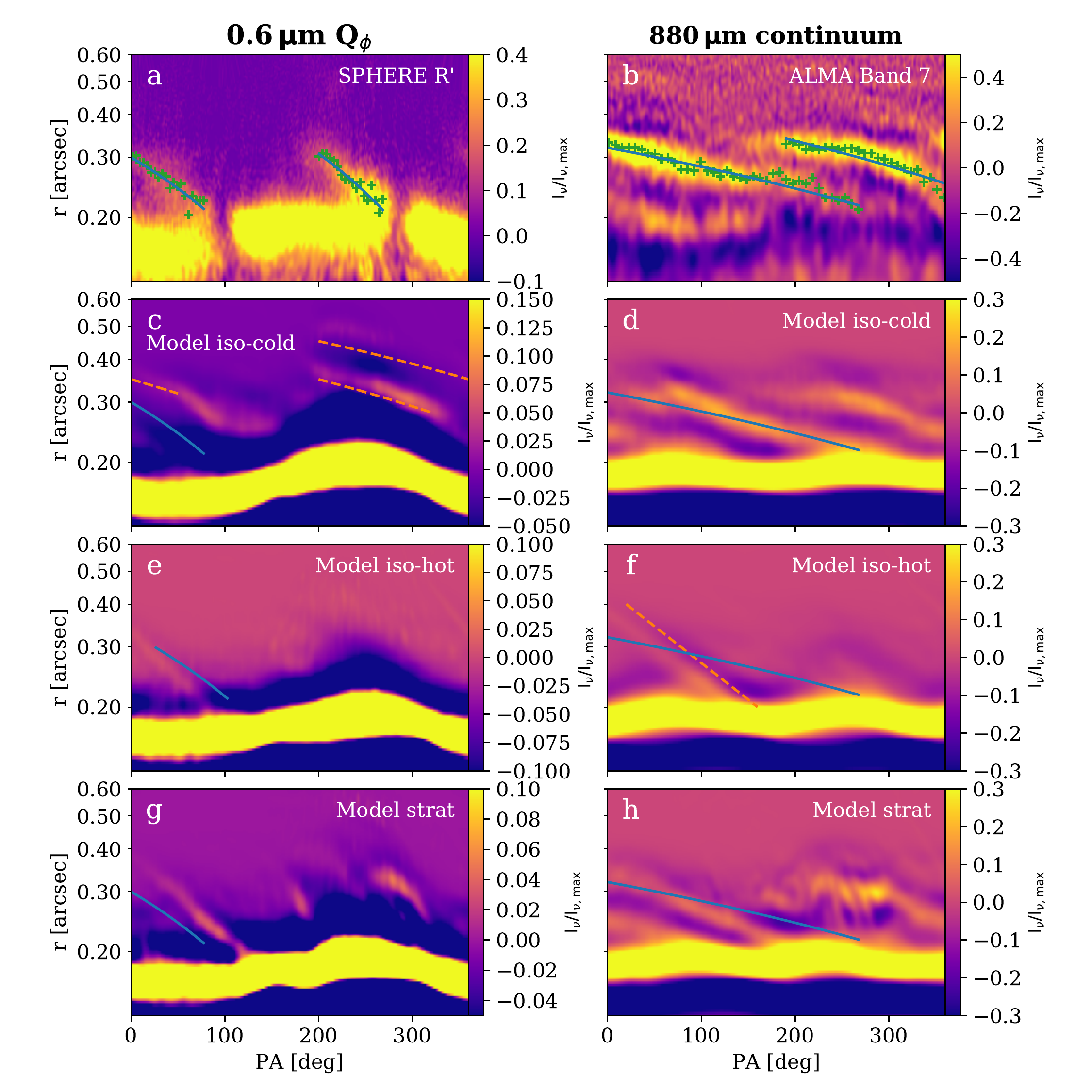}
\caption{Comparison between the data (top row) and the hydrodynamical models (rows 2-4). The images have been deprojected and high-pass filtered. The left column is for R$^\prime$ (SPHERE) and the right column for 850 $\mu$m (ALMA). The blue lines across all panels show the best-fit spiral (note we have slightly shifted it azimuthally in panel e so that it does not cover the spiral in the image), while the orange dashed line (when present) is a visual guide based on the simulated image when the simulation does not reproduce well the data. We do not plot the orange line when the simulated image reproduces well the observations. The stratified model is the only one capable of reproducing the pitch angle of the observed spiral at both wavelengths; instead the cold model only reproduces the ALMA data, while the hot model only reproduces the SPHERE data.}
\label{fig:model_comp}
\end{figure*}

\begin{figure}
    \includegraphics[width=\columnwidth]{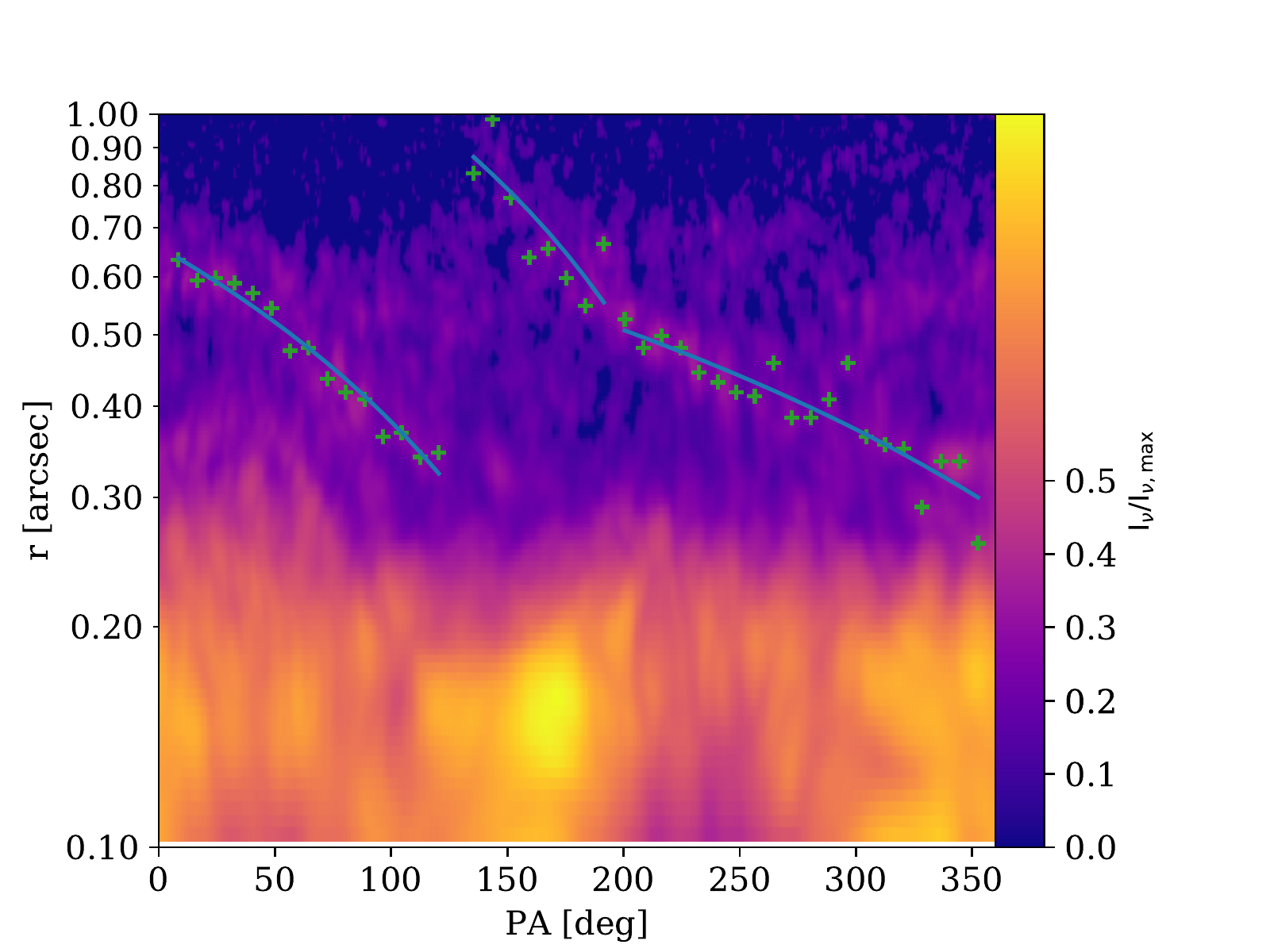}
    \caption{Peak brightness (moment 8) map of the $^{12}$CO emission in polar coordinates. Note how for $r>0.6$, the southern spiral changes pitch angle and becomes almost radial.}
    \label{fig:polar_co}
\end{figure}

We now use the geometrical parameters of the disc constrained from the modelling in Section \ref{sec:projected_velocity} to de-project the images and measure more quantitatively the pitch angles of the spirals. When deprojecting the scattered light imaging we also take into account the fact that the emission comes from a conical surface, in the same way as done by \citet{benisty_2017} (see \citealt{2016A&A...596A..70S} for details on the method employed), while we assume a razor-thin disc for the sub-mm continuum.

We show the results of this exercise in the top row of \autoref{fig:model_comp}, confirming already visually that the pitch angle in scattered light is significantly higher. To measure the pitch angle more quantitatively, we trace the position of the spiral by looking at each azimuthal angle for the radial location corresponding to the maximum in emission (inside an appropriate range to avoid picking up the bright rim). We then fit these locations with an Archmidean spiral, i.e. with equation $R=a \phi +b$, where the free parameters are $a$ and $b$, and $a$ is related to the spiral pitch angle $\mu$ as $\tan \mu = a/R$. Given the limited radial range of the continuum spirals, we do not attempt to distinguish between an Archimedean (in which the pitch angle varies with radius) and a logarithmic (constant pitch angle) spiral model. To assign uncertainties to the pitch angle, we assign a standard deviation of the radial position corresponding to the projected beam size; the beam size is also used to set the angular spacing between the tracing points, since points closer than the beam are correlated.

For scattered light, we obtain an angle of 14$\pm$2\degr for the eastern spiral and 18$\pm$3\degr for the western spiral. For the sub-mm continuum, we obtain instead a value of 4.8$\pm$0.8\degr for the eastern spiral S1 and 6.6$\pm$1.5\degr for the western spiral S2 (these values are evaluated at a projected radius of 0.26\arcsec, corresponding to the mid-point of the radial range covered by the spirals). This confirms that the scattered light spirals have a higher pitch angle than the ones in the sub-mm continuum.

We repeat the exercise also for the $^{12}$CO emission, which we show in \autoref{fig:polar_co}. The measured pitch angles in this case are 19$\pm$3\degr for the eastern spiral and 11$\pm$2\degr for the western one. This latter value is intermediate between the scattered light and sub-mm cases, possibly suggesting that (at least on this side of the disc) $^{12}$CO comes from a deeper layer than scattered light, and viceversa on the other side. For $r>0.6$\arcsec{}, the southern spiral (which connects to the companion location) changes its pitch angle significantly as it approaches the companion, becoming almost radial. To show this, we have fitted separately the tracing points of this part of the spiral, obtaining a pitch angle of 25$\pm$4\degr. This is a natural result of the interaction with a companion \citep[e.g.,][]{ogilvie_2002,rafikov_2002a} and strongly supports the scenario in which the companion is the origin of the spiral.

\section{Comparison with hydro simulations}
\label{sec:hydro_data_comparison}

\subsection{Methods}

In order to test the hypothesis that the spirals arms seen in the ALMA and SPHERE observations are launched by the stellar companion, we perform a suite of 3D hydrodynamical simulations and then post-process them with a radiative transfer code to generate mock observations, which we then compare with the data. We detail this workflow below.

\subsubsection{Hydrodynamics}

The simulations shown in this paper have been run with the code FARGO3D\footnote{\url{http://fargo.in2p3.fr/}} \citep{2016ApJS..223...11B}, which is commonly used for proto-planetary disc applications. We employ a spherical grid with 250, 80 and 
512 cells, covering the ranges [0.1, 0.6] of the binary separation, [1.22, $\pi/2$], and [0, $2\pi$] in the radial, polar and azimuthal
directions, respectively. The companion is treated like a point mass at a radius $r=1$, with a companion to star ratio $M_2/M_1=0.17$. We employ a locally isothermal equation of state, in which the sound speed $c_s$ is a function of position only, and a physical viscosity using the \citet{shakurasunyaev} prescription with a value of $\alpha=10^{-3}$. All the images shown are after 90 orbits at a radius of 30 au (roughly 13 orbits at the companion location). As shown by \citet{dong_2016}, the spiral structure settles into a steady state after 10 companion orbits, and this time is therefore enough to investigate the spiral structure.

We run three simulations. In the first two (subsequently called ``cold" and ``hot'') the isothermal $c_s$ depends only on radius: $c_s \propto r^{-1/4}$, with the normalization set such that the disc aspect ratio at the location of the companion is 0.1 for the cold simulation and 0.2 for the hot simulation. In the third simulation (``stratified"), we allow the temperature $T$ to vary as a function of the vertical coordinate. To this end we use the prescription commonly employed when fitting observations of \citet{dartois_2003} and subsequently updated by \citet{rosenfeld_2013}:
\begin{equation}
 T(r, z) = \left\{
\begin{array}{ll}
T_s + \left(T_m - T_s\right) \left[\sin \left(\frac{\pi z}{2 z_q}\right)\right]^{4} & \mbox{if $z < z_q$} \\
T_s & \mbox{if $z \ge z_q$} 
\end{array},
\right. 
    \label{eq:c_s_z}
\end{equation}
where $T_m$ is the temperature in the midplane, $T_s$ in the upper layers and $z_q$ the height of the transition between cold mid-plane and hot upper layers. We choose the temperature in the midplane to be the same as in the cold case, while we assume that the temperature in the upper layers is 4 times the value in the midplane. Therefore the temperature in the upper layers is the same as in the hot case. This value, as well as the disc aspect ratios, were chosen based on the radiative transfer model, tailored for this system, presented in \citet{benisty_2017}, assuming a mean molecular weight of 2.35 to convert from temperature to sound speed. We also assume that $z_q=3H$, where H is the disc scale-height. The initial surface density $\Sigma$ of the disc follows $\Sigma \propto r^{-1} \exp [(r/0.4)^{-4}]$, where the sharp exponential truncation is used to mimic the truncation by the companion. To assign the initial density at every point, we use the formal solution of hydrostatic equilibrium (valid for $z/r \ll 1$) in the vertical direction:
\begin{equation}
    \rho (z) = \rho_0 \frac{c_s^2 (z=0)}{c_s^2 (z)} \exp \left[-\int_0^z \frac{\Omega_K^2}{c_s^2(z')} z' \mathrm{d} z'\right],
    \label{eq:hydrostatic_eq}
\end{equation}
where $\rho_0 = \Sigma/(\sqrt{2 \pi} H)$ is the value in the midplane (because of the temperature gradient, the value of $\rho_0$ should be renormalised taking into account the actual vertical density profile, but in practise the difference is very small and we do not take it into account, see e.g. \citealt{Flock2013}), $\Omega_K$ the Keplerian velocity and for simplicity we have dropped the dependence on radius in the notation. We compute numerically the integral in \autoref{eq:hydrostatic_eq} using the trapezoidal rule. Note that, if $c_s$ does not depends on $z$, \autoref{eq:hydrostatic_eq} gives the standard Gaussian solution. Once the density has been computed, we can solve the Euler equation in the radial (cylindrical) direction assuming steady state and in this way derive the gas azimuthal velocity, taking into account the pressure gradient correction. Retaining terms of order $(z/r)^2$, in spherical coordinates the solution for the gas angular velocity reads
\begin{equation}
    \Omega^2 (R,z)= \Omega_K^2 \left(1-\frac{3 z^2}{2 r^2} \right) + \frac{1}{\rho(R,z)r} \left[ \frac{r}{R} \frac{\partial P}{ \partial R} + \frac{1}{z (1+r^2/z^2)}\frac{ \partial P}{ \partial \theta} \right],
\end{equation}
where we have used $R$ for the spherical radius and $r$ for the cylindrical radius. The partial derivatives of the pressure are evaluated on the computational grid consistently with the ZEUS \citep{1992ApJS...80..753S} algorithm (because the pressure is a zone centered quantity, the derivatives are face centered and we thus use averaging to evaluate them at the desired location).

\subsubsection{Radiative transfer}

To investigate the observational appearance of the disc perturbed by the planet, we calculate images in scattered light and sub-mm using the 3D radiative transfer code
\textsc{radmc-3d}\footnote{http://www.ita.uni-heidelberg.de/~dullemond/software/radmc-3d/}. In the radiative transfer calculation we use a 3D spherical
mesh with N$_r$=220, N$_\theta$=190, N$_\phi$=512 grid points in the radial, poloidal and azimuthal direction, respectively. The grid extent is [18,72] au for the radial grid, [0,$\pi/2$] for the poloidal and [0,2$\pi$] for the azimuthal.

We directly use the values of the gas density from the hydrodynamical simulation in the radiative transfer grid, although note that the radiative transfer grid is more extended in the poloidal coordinate than the hydrodynamical one to properly take into account photon propagation in this region. We also remove the innermost cells in the radial direction since they are affected by the boundary condition and they are not relevant to investigate the observational appearance of the spirals. We assume a gas-to-dust ratio of 100 to set the dust density. This assumption is robust for the small grains providing most of the opacity in the NIR scattered light, while, due to the larger stopping times (Stokes numbers), it is more questionable for the larger grains ($\sim$ mm-sized) providing most of the opacity at sub-mm wavelengths. Note however that here we are more interested in the spiral morphology, rather in the amplitude of the spiral features. The amplitude of the spiral might be smaller in the large grains than in the gas, depending on their Stokes number; however, the grains respond to the spiral structure in the gas and it is therefore plausible that they produce the same morphology. We caution however that, to the best of our knowledge, in the literature there is no study focusing on the dependence of planetary spiral pitch angles with the Stokes number of the grains. We use 10 logarithmically spaced grain size bins between 0.1\,$\mu$m and 1\,mm and assume that the dust grain size distribution follows $dN/da \propto a^{-3.5}$ \citep{1977ApJ...217..425M}. To normalise the mass of the disc, we assume a total (gas) disc mass of $7 \times 10^{-4} \, M_\odot$, consistent with the values derived by \citet{Collins2009}.

The mass absorption coefficients of the dust grains are calculated 
with Mie-theory using the optical constants of astronomical silicates \citep{Weingartner2001}. The radiation field of the central star is modelled
with blackbody emission and the star is assumed to have M$_\star$=1.66\,M$_\odot$, T$_{\rm eff}$ = 7400\,K, R$_\star$=1.73\,R$_\odot$. 

As a first step, we calculate the temperature of the dust with a thermal Monte Carlo simulation, then we calculate images at 1.65\,$\mu$m and 880\,$\mu$m taking the disc inclination to be the one derived in Section \ref{sec:projected_velocity}. We use $10^7$ photons both for the thermal Monte Carlo simulations and for the image calculations.

\subsection{Results}
\label{sec:results}

We show in \autoref{fig:model_comp} a comparison between the data and the three hydrodynamical simulations we have run. Data and models in R$^\prime$ (SPHERE) are in the left column and at 850 $\mu$m (ALMA) in the right column. We plot all the images in polar coordinates; the deprojection also takes into account the fact that the emission comes from a cone for the scattered light case (with the parameters of Section \ref{sec:geometric_model}). The images have also been enhanced by a high pass filter (see Section \ref{subsec:cont_image}). Note that the edge of the cavity in scattered light does not deproject into a circular ring because at certain position angles one can see also the bottom side of the disc.

It can be seen how reproducing the spiral in scattered light requires a high temperature: the cold model produces a spiral that is too closed (i.e., too low pitch-angle) in comparison to the observations. On the other hand, the predictions for the hot and stratified model (these two models have the same temperature in the disc upper layers) are consistent with the data. A similar result has been found also by \citet{dong_2016}, who also required a high disc temperature to reproduce the high opening angle of the spirals in the SPHERE image.

In the same way, reproducing the spiral in the sub-mm continuum requires a cold temperature: the hot model produces a spiral that is too open in comparison to the observations, while the cold and stratified models are successful in reproducing the observed pitch angle.

The comparison clearly shows that the stratified model is the only one capable of reproducing the pitch angles of the scattered light and sub-mm observations \textit{at the same time}. This model has a realistic vertical temperature structure, which is commonly found in passively irradiated proto-planetary discs \citep{Calvet1991,chiang_1997,Dalessio1998,Dullemond2001}. Therefore, the data presented in this paper strongly support the theoretical prediction formulated by \citet{juhasz_2018} that the pitch angle of the spirals varies not only as a function of radius, but also as a function of height above the midplane due to the dependence of the pitch angle on the local sound speed. At the same time, the fact that we can correctly account for the observed spiral morphology lends further credence to the scenario in which the origin of the spiral arms is the nearby M-dwarf companion, as originally proposed by \citet{dong_2016}.

\section{Discussion}
\label{sec:discussion}

\subsection{On the extent of the CO disc}
\label{sec:extent_co}

\begin{figure}
    \centering
    \includegraphics[width=\columnwidth]{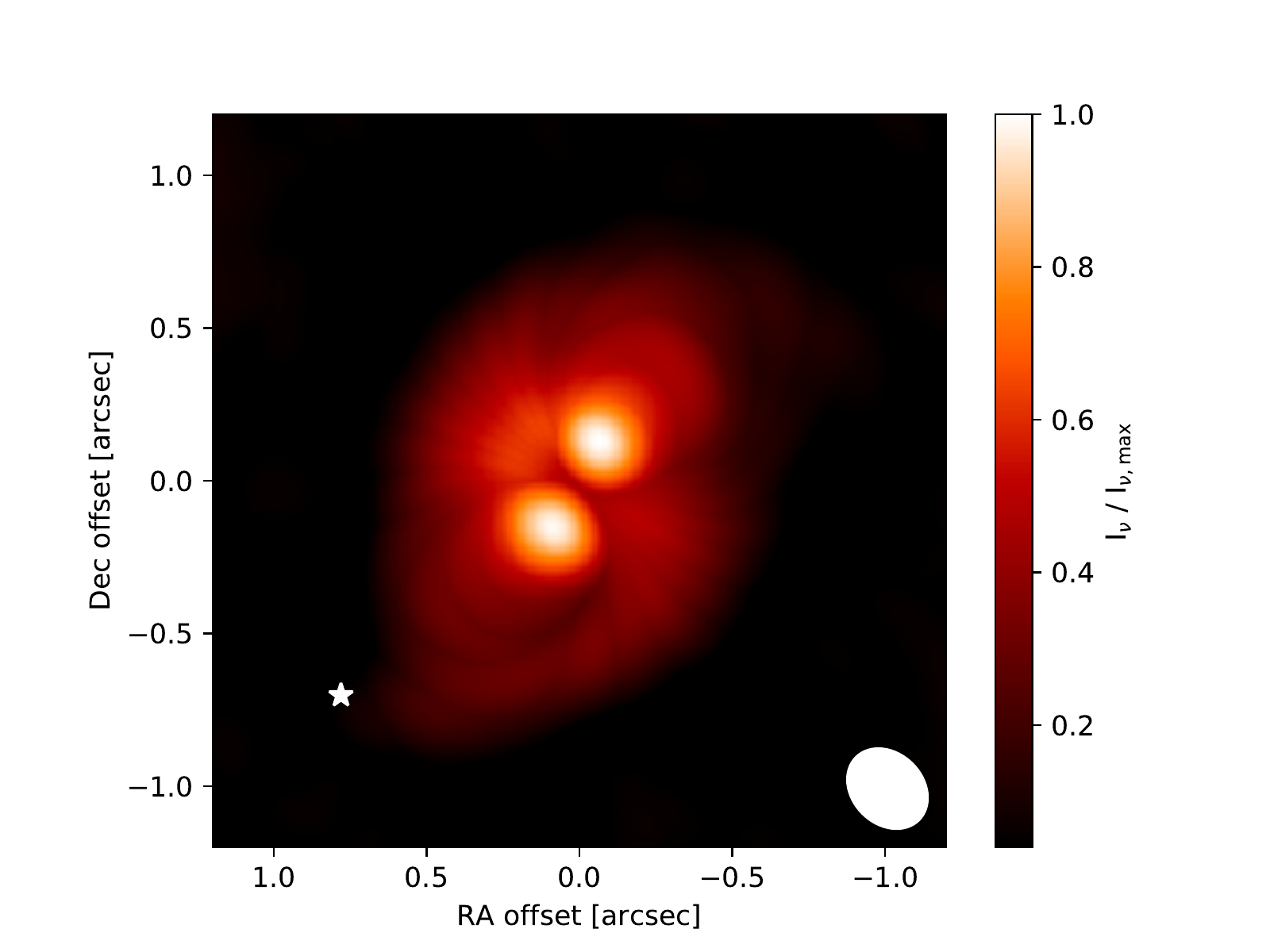}
    \caption{Peak intensity (8th moment) map computed from the data smoothed to the  resolution of  the Band 6 data \citep{vanderplas2019}. At low resolution, the spirals cannot be distinguished from the disc and this creates the impression of a much larger disc, that extends almost up to the companion.}
    \label{fig:12co_smoothed}
\end{figure}

In recently published \citep{vanderplas2019} ALMA Band 6 observations, the CO disc is significantly more extended than in our observations: emission can be traced almost up to the location of the stellar companion (1.05\arcsec{} projected separation). This is puzzling since co-planar companions are supposed to truncate circumstellar discs at roughly one third of the separation \citep{artymowicz1994}. This led \citet{vanderplas2019} to dispute that the companion is co-planar with the disc and suggest that its orbit might lie on another plane.

The resolution of the Band 6 data in $^{12}$CO J=2-1 is 0.29 $\times$ 0.23\arcsec{}, a factor of $\sim$5 lower than the $^{12}$CO J=3-2 Band 7 data we present in this paper. 
In the Band 7 data, as we have highlighted in Section \ref{subsec:co_image}, the CO disc extends only up to $\sim$0.3\arcsec{}, which is in broad agreement with the expected truncation radius. Outside this radius, there is no full disc, but only the two spiral arms S3 and S4, raising the possibility that such structures were misinterpreted as a axisymmetric disc in the low resolution dataset. There are however two caveats about our data: at such a high resolution, the surface brightness sensitivity is significantly lower; in addition the maximum recoverable scale is 0.6\arcsec{}, smaller than the separation of the companion (although, note that in practise the requirement on the maximum recoverable scale is not so severe because the gas emission in each single channel comes from only a portion of the disc). In principle, it is therefore possible that we are missing emission from an extended disc. 

To study whether there is indeed an extended disc, we have lowered the resolution of our CO map, using the task \texttt{imsmooth} in CASA on the individual channel maps. We then recomputed the moment maps from the low resolution channel maps. In this discussion we consider only the moment 8 (peak intensity) map: the moment 0 map in \citealt{vanderplas2019} is strongly centrally peaked and it is difficult to see the extended emission, while the moment 1 map does not contain any additional information regarding the extent of the disc.

We plot the result of this exercise in \autoref{fig:12co_smoothed}.  This map is remarkably similar to the one obtained with the Band 6 data  \citep{vanderplas2019} (see their Figure 2). While this exercise does not formally prove that we are not missing flux do the long baselines, it does prove that, if we are missing some flux, the effect is too small to affect the morphology of the emission. Therefore, the two caveats we listed above do not affect our conclusion: the large extent of the CO disc is likely an artefact of the low resolution, which does not allow one to distinguish the spirals from the disc. Note how the southern spiral can still be seen in the low resolution moment 8 map, as well as in the map of \citet{vanderplas2019}; however its identification would be dubious without the support of the high resolution data we present in this paper.

Therefore, the higher resolution ALMA data we present in this paper is compatible with an orbit of the companion aligned with the plane of the disc and with the companion truncating the disc. The orbit could also be misaligned, but the ALMA data does not favour a specific scenario. Further hydrodynamical studies, beyond the scope of this paper, focusing on the truncation radius of the CO disc could provide further constraints regarding the orbit of the companion. We note that the analysis of the astrometry of the companion is indeed compatible with a broad range of values\footnote{Although there is a shallow maximum at 60\degr, the posterior is essentially uniform between 10-80\degr.} for the relative inclination between the disc and the companion orbit (see fig. 9d in \citealt{vanderplas2019}). Given that the existing astrometry goes as far back as 2003, following up the orbit of the companion for many years (probably at least a decade) will be needed to improve significantly the constraints on the inclination from astrometry.

\subsection{On the origin of the spirals and model limitations} \label{sec:origin_spirals}

In the context of the current debate about the origin of observed spiral arms in proto-planetary discs, the data presented in this paper contain two pieces of evidence that strongly point to the companion as responsible for the spirals, at least for this object. The first one is the fact that the southern CO spiral points to the location of the companion. The second is the difference in pitch angles between midplane (ALMA continuum) and upper layers (NIR scattered light).

While the dynamical scenario is mainly successful, it should be noted that it does not fully account for the observed morphology of the spirals. In particular, the spirals in scattered light observations are symmetrical, while in the hydrodynamical simulations (see \autoref{fig:model_comp}) we find that one spiral is stronger than the other. While this problem is particularly severe for the hot model, which does not reproduce the pitch angle of the continuum spiral, it is still present in the stratified model, the only one capable of reproducing at the same time the spiral pitch angles in the mid-plane and in the upper layers. This problem is present also in the hydro simulations presented by \citet{2018ApJ...854..130W} (see their Figure 7). Reconciling this discrepancy might require considering a non-vanishing relative inclination between the companion orbit and the disc, while for simplicity the simulations presented in this paper have considered a non-inclined orbit. The grain scattering phase function is also another factor that might change the brightness of the spiral arms since it strongly determines the amount of light scattered along the line of sight.

In addition, the scattered light image also shows two dark spots in the central ring, that \citet{benisty_2017} interpreted as due to shadows cast by a misaligned inner disc, likely to be on spatial scales smaller than those we resolve in our observations. In the companion scenario, it remains unexplained why the scattered light shadows lie very close to where the scattered light spirals detach from the inner ring. To explain this coincidence, \citet{Montesinos2016} proposed that the shadows are actually the \textit{cause} of the spirals and confirmed through hydrodynamical simulations that the lower pressure at the shadow locations produces a variable azimuthal acceleration that in time develops into spiral density waves. However, there are no strong, obvious sub-mm counterparts of the NIR shadows (see appendix \ref{sec:shadows}), implying that in this source the shadows do not cause a significant temperature, and therefore pressure, drop in the mid-plane. Following the framework developed by \citet{Casassus2019}, this can be explained as due to the effect of radiation smoothing temperature differences, implying that the material is optically thin to radiative diffusion (i.e., with respect to the Rosseland mean opacity). We cannot assess quantitatively whether this condition is verified because we do not have information on the grain size and therefore the Rosseland opacity; we note however that the sub-mm continuum emission is largely optically thin: the maximum brightness temperature across the image is 18 K, attained at 0.3" from the star in the North-East(from a simple estimate using the luminosity of the star, e.g. \citealt{Dullemond2018}, we would expect a temperature of 30 K at that location), but most of the emission is fainter than that (see left panel of \autoref{fig:continuum_image}). Therefore, it is plausible that the disc is optically thin to radiative diffusion as long as the Rosseland mean opacity is not much higher than the sub-mm opacity. In the scenario in which the shadows launch the spirals there is also no reason why the CO spiral should point to the location of the companion. Moreover, follow up simulations including also dust dynamics \citep{Cuello2019} indicated that there should be no observable sub-mm continuum spiral produced by this mechanism, in contrast with our data. Given that all these facts rule out launching by shadows as origin for the spirals, it is possible that the special shadow location is just a lucky coincidence. Future observations will be able to tell if the spirals rotate with the companion or with the shadows, though this test might require very long timespans due to the long orbital timescale of the companion.

Another limitation of our modelling is that we have not studied the formation of a circum-secondary disc. In principle we could expect that some of the material in the spiral arms should circularise around the secondary, forming another disc (see e.g. the simulations presented by \citealt{vanderplas2019}); however there is no evidence for this in the CO emission. We speculate that this disc might accrete very rapidly and therefore be short lived, possibly due to the effect of tidal truncation coupled with viscosity \citep{RosottiClarke2018}, but we note that this should be the subject of a future study.

Finally, the last limitation to highlight in our modelling is that
we have assumed that the spirals in the ALMA continuum image
trace the same morphology as the spirals in the midplane gas. While
this is plausible, this is currently untested and has not been yet the
subject of a dedicated study. Future work will establish under which
conditions the assumption holds.


\subsection{Comparison with other discs with spirals}
To the best of our knowledge, this is the first source  where there are {\it two} sub-mm continuum counterparts to spirals observed in scattered light. The fact that spirals are seen in all tracers confirms that they are real perturbations in surface density, in this case launched by the stellar companion. For what concerns other sources, most other discs showing spirals in scattered light, when they have been imaged in sub-mm continuum \citep[e.g.,][]{Kraus2017,Cazzoletti2018}, show structures like vortices and crescents rather than spirals. MWC758 \citep{Dong2018} is notable because, on top of vortices, also shows a spiral in the sub-mm continuum. Note however that only one spiral arm is visible, while the scattered light signal \citep{2013ApJ...762...48G,benisty_2015} shows two arms and is very similar in morphology to HD100453. Recent hydrodynamic simulations \citep{Baruteau2019} suggest that the morphology of these objects with vortices could be explained by two massive planets rather than a stellar companion. These planets trigger vortices trapping the large mm grains seen in the sub-mm, possibly explaining the reason for the different morphology between sub-mm and scattered light. The simulations did not target specifically reproducing the single spiral arm observed in MWC758, although there is some hint that reproducing it is sensitive to the amount of small grains.

On the other hand, there is now a small sample of sources with detected spirals in sub-mm continuum. Some of them are in known stellar multiple systems \citep{2018ApJ...869L..44K}; in this case it is likely that the stellar companion is responsible for the spirals. Elias 2-27 \citep{perez_2016} is instead a good candidate for an origin due to gravitational instability \citep{meru_2017,2018MNRAS.477.1004H}, though the possibility of an external companion has not been completely ruled out. In the other two cases \citep{huang2018b} the launching mechanism has not been clearly identified. Among all of these, IM Lup is the only one with published observations in scattered light \citep{Avenhaus2018}. It is important to note that in the single case of IM Lup the scattered light image, while showing azimuthally symmetric structure, does not show any sign of a spiral.

Summarising, it is clear that the morphology can vary significantly from source to source, especially when combining multi-wavelength data (sub-mm and scattered light) in the limited cases in which this is possible. This richness in morphology probably points to different formation mechanisms operating in discs, rather than a single universal process. 


\subsection{What is causing the inner cavity?}
In this paper we focused on the two prominent spiral arms. However, as already discussed the source is a known transition disc, with a very well defined ring at 0.2\arcsec{} from the star. It is clear that this structure cannot be due to the external companion and another process must be invoked. There is a large literature (see \citealt{Espaillat2014,2017RSOS....470114E} for reviews of the topic) about the mechanisms causing transition discs and here we only briefly summarise them. The leading interpretation is planet-disc interaction \citep[e.g.,][]{Rice2006,Pinilla2012}, which would require postulating the presence of a planet causing the ring. The planet mass should be higher than the canonical ``pebble isolation mass'' \citep{Lambrechts2014,Rosotti2016} to produce a ring in the sub-mm continuum. Depending on the value of the disc viscosity \citep[e.g.,][]{Bitsch2018,Zhang2018}, this could be possible with a sub-Jupiter mass planet, well below the existing detection limits of direct imaging. 

According to the predictions (see their Figures 6 and 8) of \citet{Facchini2018}, the putative planet cannot be more massive than Jupiter, or it would produce a detectable gap in $^{12}$CO, in contrast with our observations. In the case of PDS 70 \citep{2019arXiv190207639K}, the directly imaged companion produces a clear gap in $^{12}$CO, likely indicating that the putative planet in HD100453 must have a lower mass. \citet{vanderplas2019} recently suggested that this putative planet is also responsible for misaligning the inner disc (which produces the shadows in scattered light), following the suggestion of \citet{2017MNRAS.469.2834O} that this can happen due to a secular resonance between the nodal precession of the inner disc and the precession of the putative companion. Given the constraints on the planet mass, it is unclear whether this is indeed possible since the mechanism requires masses of at least 0.01 $M_\odot$. It could be that an additional companion at smaller spatial scales (or a different mechanism from planets) is required to explain the misaligned inner disc.

The presence of CO emission well inside the continuum ring tends to rule out photo-evaporation \citep[e.g.,][]{Owen2011} as a possible formation mechanism of the ring. On the other hand, the lack of detected accretion onto the star \citep{Collins2009} could mean that we are observing this source at a particular moment in time while the hole opened by photo-evaporation is still expanding and the inner disc has not completely dissipated, possibly reconciling a photo-evaporative origin with these observations. 


\section{Conclusions}
\label{sec:conclusions}

In this paper we have presented high-resolution (0.03\arcsec{}) continuum and $^{12}$CO J=3-2 maps of the proto-planetary disc around HD100453. Our main results are as follows:

\begin{itemize}
    \item The source shows two, almost symmetrical spiral arms both in the continuum and in the CO emission. The continuum spirals have a relatively narrow radial range (0.2-0.35\arcsec{}), while the gas spirals start from outside the continuum spirals (0.3\arcsec{}) and extend for much further (up to 1\arcsec{}).
    \item The southern gas spiral connects to the companion location, implying that the spirals are the result of the tidal interaction between the disc and the companion.
    \item The intrinsic pitch angle of the spirals in the continuum (6 \degr{}) is significantly lower than in the SPHERE scattered light images (19 \degr{}). This confirms the theoretical prediction of \citet{juhasz_2018} and can be explained as due to the different temperatures between the cold disc midplane and the hot upper layers. This difference also further reinforces the hypothesis that the spiral pattern is due to the interaction with the companion.
    \item Through 3D hydrodynamical simulations with a stratified thermal structure, we show that the difference in pitch angles between sub-mm and scattered light can be accounted for quantitatively. Although two spirals are present in the simulation, they are not symmetrical as in the observations (particularly for the scattered light case), an issue that was already present in the simulations of \citet{2018ApJ...854..130W}. Solving this discrepancy will require exploring a possible misaligment between the disc and the companion orbit, as well as exploring the grain scattering phase function. 
    \item The high spatial resolution of our data allows us to conclude that the CO disc extends only up to 0.3\arcsec{}, which is roughly one third of the separation from the companion. Outside this radius, there is no emission from a disc but only two spiral arms. This solves the apparent discrepancy between the companion location and the disc truncation radius reported by previous, low resolution observations \citep{vanderplas2019}. It also implies  that the orbit of the companion is compatible (though this is not necessarily the case) with lying in the same plane as the disc.
\end{itemize}


\section*{Acknowledgements}
We thank the referee, Simon Casassus, for a careful reading of our manuscript and the constructive criticism. This paper makes use of the following ALMA data: ADS/JAO.ALMA\#2017.1.01424.S. ALMA is a partnership of ESO (representing its member states), NSF (USA) and NINS (Japan), together with NRC (Canada), MOST and ASIAA (Taiwan), and KASI (Republic of Korea), in cooperation with the Republic of Chile. The Joint ALMA Observatory is operated by ESO, AUI/NRAO and NAOJ. This work has been supported by the DISCSIM project, grant agreement 341137 funded by the European Research Council under ERC-2013-ADG. G.R. acknowledges support from the Netherlands Organisation for Scientific Research (NWO, program number 016.Veni.192.233). This work was performed using the Cambridge Service for Data Driven Discovery (CSD3), part of which is operated by the University of Cambridge Research Computing on behalf of the STFC DiRAC HPC Facility (www.dirac.ac.uk). The DiRAC component of CSD3 was funded by BEIS capital funding via STFC capital grants ST/P002307/1 and ST/R002452/1 and STFC operations grant ST/R00689X/1. DiRAC is part of the National e-Infrastructure. M.B. acknowledges funding from ANR of France under contract number ANR-16-CE31-0013 (Planet Forming disks). This
project has received funding from the European Union’s Horizon 2020
research and innovation programme under the Marie Skłodowska-Curie grant
greement No 823823 (DUSTBUSTERS). C.D. acknowledges funding from the Netherlands Organisation for Scientific Research (NWO) TOP-1 grant as part of the research programme “Herbig Ae/Be stars, Rosetta stones for understanding the formation of planetary systems”, project number 614.001.552. T.S. acknowledges the support from the ETH Zurich Postdoctoral Fellowship Program.


\bibliographystyle{mnras}
\bibliography{ms}

\begin{thebibliography}{}
\makeatletter
\relax
\def\mn@urlcharsother{\let\do\@makeother \do\$\do\&\do\#\do\^\do\_\do\%\do\~}
\def\mn@doi{\begingroup\mn@urlcharsother \@ifnextchar [ {\mn@doi@}
  {\mn@doi@[]}}
\def\mn@doi@[#1]#2{\def\@tempa{#1}\ifx\@tempa\@empty \href
  {http://dx.doi.org/#2} {doi:#2}\else \href {http://dx.doi.org/#2} {#1}\fi
  \endgroup}
\def\mn@eprint#1#2{\mn@eprint@#1:#2::\@nil}
\def\mn@eprint@arXiv#1{\href {http://arxiv.org/abs/#1} {{\tt arXiv:#1}}}
\def\mn@eprint@dblp#1{\href {http://dblp.uni-trier.de/rec/bibtex/#1.xml}
  {dblp:#1}}
\def\mn@eprint@#1:#2:#3:#4\@nil{\def\@tempa {#1}\def\@tempb {#2}\def\@tempc
  {#3}\ifx \@tempc \@empty \let \@tempc \@tempb \let \@tempb \@tempa \fi \ifx
  \@tempb \@empty \def\@tempb {arXiv}\fi \@ifundefined
  {mn@eprint@\@tempb}{\@tempb:\@tempc}{\expandafter \expandafter \csname
  mn@eprint@\@tempb\endcsname \expandafter{\@tempc}}}

\bibitem[\protect\citeauthoryear{{ALMA Partnership} et~al.,}{{ALMA Partnership}
  et~al.}{2015}]{alma-partnership_2015}
{ALMA Partnership} et~al., 2015, \mn@doi [\apjl] {10.1088/2041-8205/808/1/L3},
  \href {http://adsabs.harvard.edu/abs/2015ApJ...808L...3A} {808, L3}

\bibitem[\protect\citeauthoryear{{Artymowicz} \& {Lubow}}{{Artymowicz} \&
  {Lubow}}{1994}]{artymowicz1994}
{Artymowicz} P.,  {Lubow} S.~H.,  1994, \mn@doi [\apj] {10.1086/173679}, \href
  {https://ui.adsabs.harvard.edu/abs/1994ApJ...421..651A} {421, 651}

\bibitem[\protect\citeauthoryear{{Avenhaus} et~al.,}{{Avenhaus}
  et~al.}{2018}]{Avenhaus2018}
{Avenhaus} H.,  et~al., 2018, \mn@doi [\apj] {10.3847/1538-4357/aab846}, \href
  {https://ui.adsabs.harvard.edu/abs/2018ApJ...863...44A} {863, 44}

\bibitem[\protect\citeauthoryear{{Bae} \& {Zhu}}{{Bae} \&
  {Zhu}}{2018}]{2018ApJ...859..119B}
{Bae} J.,  {Zhu} Z.,  2018, \mn@doi [\apj] {10.3847/1538-4357/aabf93}, \href
  {https://ui.adsabs.harvard.edu/#abs/2018ApJ...859..119B} {859, 119}

\bibitem[\protect\citeauthoryear{{Baruteau} et~al.,}{{Baruteau}
  et~al.}{2019}]{Baruteau2019}
{Baruteau} C.,  et~al., 2019, \mn@doi [\mnras] {10.1093/mnras/stz802}, \href
  {https://ui.adsabs.harvard.edu/abs/2019MNRAS.486..304B} {486, 304}

\bibitem[\protect\citeauthoryear{{Benisty} et~al.,}{{Benisty}
  et~al.}{2015}]{benisty_2015}
{Benisty} M.,  et~al., 2015, \mn@doi [\aap] {10.1051/0004-6361/201526011},
  \href {http://adsabs.harvard.edu/abs/2015A%26A...578L...6B} {578, L6}

\bibitem[\protect\citeauthoryear{{Benisty} et~al.,}{{Benisty}
  et~al.}{2017}]{benisty_2017}
{Benisty} M.,  et~al., 2017, \mn@doi [\aap] {10.1051/0004-6361/201629798},
  \href {http://adsabs.harvard.edu/abs/2017A%26A...597A..42B} {597, A42}

\bibitem[\protect\citeauthoryear{{Ben{\'\i}tez-Llambay} \&
  {Masset}}{{Ben{\'\i}tez-Llambay} \& {Masset}}{2016}]{2016ApJS..223...11B}
{Ben{\'\i}tez-Llambay} P.,  {Masset} F.~S.,  2016, \mn@doi [The Astrophysical
  Journal Supplement Series] {10.3847/0067-0049/223/1/11}, \href
  {https://ui.adsabs.harvard.edu/#abs/2016ApJS..223...11B} {223, 11}

\bibitem[\protect\citeauthoryear{{Bitsch}, {Morbidelli}, {Johansen}, {Lega},
  {Lambrechts}  \& {Crida}}{{Bitsch} et~al.}{2018}]{Bitsch2018}
{Bitsch} B.,  {Morbidelli} A.,  {Johansen} A.,  {Lega} E.,  {Lambrechts} M.,
  {Crida} A.,  2018, \mn@doi [\aap] {10.1051/0004-6361/201731931}, \href
  {https://ui.adsabs.harvard.edu/abs/2018A&A...612A..30B} {612, A30}

\bibitem[\protect\citeauthoryear{{Boehler} et~al.,}{{Boehler}
  et~al.}{2018}]{Boehler2018}
{Boehler} Y.,  et~al., 2018, \mn@doi [\apj] {10.3847/1538-4357/aaa19c}, \href
  {https://ui.adsabs.harvard.edu/#abs/2018ApJ...853..162B} {853, 162}

\bibitem[\protect\citeauthoryear{{Boss}}{{Boss}}{2002}]{2002ApJ...567L.149B}
{Boss} A.~P.,  2002, \mn@doi [\apj] {10.1086/340108}, \href
  {https://ui.adsabs.harvard.edu/abs/2002ApJ...567L.149B} {567, L149}

\bibitem[\protect\citeauthoryear{{Calvet}, {Patino}, {Magris}  \&
  {D'Alessio}}{{Calvet} et~al.}{1991}]{Calvet1991}
{Calvet} N.,  {Patino} A.,  {Magris} G.~C.,   {D'Alessio} P.,  1991, \mn@doi
  [\apj] {10.1086/170618}, \href
  {https://ui.adsabs.harvard.edu/abs/1991ApJ...380..617C} {380, 617}

\bibitem[\protect\citeauthoryear{{Casassus} \& {Perez}}{{Casassus} \&
  {Perez}}{2019}]{2019arXiv190606302C}
{Casassus} S.,  {Perez} S.,  2019, arXiv e-prints, \href
  {https://ui.adsabs.harvard.edu/abs/2019arXiv190606302C} {p. arXiv:1906.06302}

\bibitem[\protect\citeauthoryear{{Casassus} et~al.,}{{Casassus}
  et~al.}{2013}]{Casassus2013}
{Casassus} S.,  et~al., 2013, \mn@doi [\nat] {10.1038/nature11769}, \href
  {https://ui.adsabs.harvard.edu/abs/2013Natur.493..191C} {493, 191}

\bibitem[\protect\citeauthoryear{{Casassus}, {P{\'e}rez}, {Osses}  \&
  {Marino}}{{Casassus} et~al.}{2019}]{Casassus2019}
{Casassus} S.,  {P{\'e}rez} S.,  {Osses} A.,   {Marino} S.,  2019, \mn@doi
  [\mnras] {10.1093/mnrasl/slz059}, \href
  {https://ui.adsabs.harvard.edu/abs/2019MNRAS.486L..58C} {486, L58}

\bibitem[\protect\citeauthoryear{{Cazzoletti} et~al.,}{{Cazzoletti}
  et~al.}{2018}]{Cazzoletti2018}
{Cazzoletti} P.,  et~al., 2018, \mn@doi [\aap] {10.1051/0004-6361/201834006},
  \href {https://ui.adsabs.harvard.edu/abs/2018A&A...619A.161C} {619, A161}

\bibitem[\protect\citeauthoryear{{Chen}, {Henning}, {van Boekel}  \&
  {Grady}}{{Chen} et~al.}{2006}]{Chen2006}
{Chen} X.~P.,  {Henning} T.,  {van Boekel} R.,   {Grady} C.~A.,  2006, \mn@doi
  [\aap] {10.1051/0004-6361:20054122}, \href
  {https://ui.adsabs.harvard.edu/#abs/2006A&A...445..331C} {445, 331}

\bibitem[\protect\citeauthoryear{{Chiang} \& {Goldreich}}{{Chiang} \&
  {Goldreich}}{1997}]{chiang_1997}
{Chiang} E.~I.,  {Goldreich} P.,  1997, \mn@doi [\apj] {10.1086/304869}, \href
  {http://adsabs.harvard.edu/abs/1997ApJ...490..368C} {490, 368}

\bibitem[\protect\citeauthoryear{{Christiaens}, {Casassus}, {Perez}, {van der
  Plas}  \& {M{\'e}nard}}{{Christiaens} et~al.}{2014}]{Christiaens2014}
{Christiaens} V.,  {Casassus} S.,  {Perez} S.,  {van der Plas} G.,
  {M{\'e}nard} F.,  2014, \mn@doi [\apjl] {10.1088/2041-8205/785/1/L12}, \href
  {https://ui.adsabs.harvard.edu/abs/2014ApJ...785L..12C} {785, L12}

\bibitem[\protect\citeauthoryear{{Clarke} et~al.,}{{Clarke}
  et~al.}{2018}]{2018ApJ...866L...6C}
{Clarke} C.~J.,  et~al., 2018, \mn@doi [\apj] {10.3847/2041-8213/aae36b}, \href
  {https://ui.adsabs.harvard.edu/\#abs/2018ApJ...866L...6C} {866, L6}

\bibitem[\protect\citeauthoryear{{Collins} et~al.,}{{Collins}
  et~al.}{2009}]{Collins2009}
{Collins} K.~A.,  et~al., 2009, \mn@doi [\apj] {10.1088/0004-637X/697/1/557},
  \href {http://adsabs.harvard.edu/abs/2009ApJ...697..557C} {697, 557}

\bibitem[\protect\citeauthoryear{{Cossins}, {Lodato}  \& {Clarke}}{{Cossins}
  et~al.}{2009}]{cossins_2009}
{Cossins} P.,  {Lodato} G.,   {Clarke} C.~J.,  2009, \mn@doi [\mnras]
  {10.1111/j.1365-2966.2008.14275.x}, \href
  {http://adsabs.harvard.edu/abs/2009MNRAS.393.1157C} {393, 1157}

\bibitem[\protect\citeauthoryear{{Cossins}, {Lodato}  \& {Testi}}{{Cossins}
  et~al.}{2010}]{Cossins2010}
{Cossins} P.,  {Lodato} G.,   {Testi} L.,  2010, \mn@doi [\mnras]
  {10.1111/j.1365-2966.2010.16934.x}, \href
  {https://ui.adsabs.harvard.edu/#abs/2010MNRAS.407..181C} {407, 181}

\bibitem[\protect\citeauthoryear{{Cuello}, {Montesinos}, {Stammler}, {Louvet}
  \& {Cuadra}}{{Cuello} et~al.}{2019}]{Cuello2019}
{Cuello} N.,  {Montesinos} M.,  {Stammler} S.~M.,  {Louvet} F.,   {Cuadra} J.,
  2019, \mn@doi [\aap] {10.1051/0004-6361/201731732}, \href
  {https://ui.adsabs.harvard.edu/abs/2019A&A...622A..43C} {622, A43}

\bibitem[\protect\citeauthoryear{{D'Alessio}, {Cant{\"o}}, {Calvet}  \&
  {Lizano}}{{D'Alessio} et~al.}{1998}]{Dalessio1998}
{D'Alessio} P.,  {Cant{\"o}} J.,  {Calvet} N.,   {Lizano} S.,  1998, \mn@doi
  [\apj] {10.1086/305702}, \href
  {https://ui.adsabs.harvard.edu/abs/1998ApJ...500..411D} {500, 411}

\bibitem[\protect\citeauthoryear{{Dartois}, {Dutrey}  \&
  {Guilloteau}}{{Dartois} et~al.}{2003}]{dartois_2003}
{Dartois} E.,  {Dutrey} A.,   {Guilloteau} S.,  2003, \mn@doi [\aap]
  {10.1051/0004-6361:20021638}, \href
  {http://adsabs.harvard.edu/abs/2003A%26A...399..773D} {399, 773}

\bibitem[\protect\citeauthoryear{{Dipierro}, {Lodato}, {Testi}  \& {de Gregorio
  Monsalvo}}{{Dipierro} et~al.}{2014}]{2014MNRAS.444.1919D}
{Dipierro} G.,  {Lodato} G.,  {Testi} L.,   {de Gregorio Monsalvo} I.,  2014,
  \mn@doi [\mnras] {10.1093/mnras/stu1584}, \href
  {https://ui.adsabs.harvard.edu/#abs/2014MNRAS.444.1919D} {444, 1919}

\bibitem[\protect\citeauthoryear{{Dominik}, {Dullemond}, {Waters}  \&
  {Walch}}{{Dominik} et~al.}{2003}]{Dominik2003}
{Dominik} C.,  {Dullemond} C.~P.,  {Waters} L.~B.~F.~M.,   {Walch} S.,  2003,
  \mn@doi [\aap] {10.1051/0004-6361:20021629}, \href
  {http://adsabs.harvard.edu/abs/2003A%26A...398..607D} {398, 607}

\bibitem[\protect\citeauthoryear{{Dong}, {Zhu}, {Rafikov}  \& {Stone}}{{Dong}
  et~al.}{2015}]{dong_2015a}
{Dong} R.,  {Zhu} Z.,  {Rafikov} R.~R.,   {Stone} J.~M.,  2015, \mn@doi [\apjl]
  {10.1088/2041-8205/809/1/L5}, \href
  {http://adsabs.harvard.edu/abs/2015ApJ...809L...5D} {809, L5}

\bibitem[\protect\citeauthoryear{{Dong}, {Zhu}, {Fung}, {Rafikov}, {Chiang}  \&
  {Wagner}}{{Dong} et~al.}{2016}]{dong_2016}
{Dong} R.,  {Zhu} Z.,  {Fung} J.,  {Rafikov} R.,  {Chiang} E.,   {Wagner} K.,
  2016, \mn@doi [\apjl] {10.3847/2041-8205/816/1/L12}, \href
  {http://adsabs.harvard.edu/abs/2016ApJ...816L..12D} {816, L12}

\bibitem[\protect\citeauthoryear{{Dong} et~al.,}{{Dong}
  et~al.}{2018}]{Dong2018}
{Dong} R.,  et~al., 2018, preprint, \href
  {https://ui.adsabs.harvard.edu/#abs/2018arXiv180512141D} {p.
  arXiv:1805.12141} (\mn@eprint {arXiv} {1805.12141})

\bibitem[\protect\citeauthoryear{{Dullemond}, {Dominik}  \&
  {Natta}}{{Dullemond} et~al.}{2001}]{Dullemond2001}
{Dullemond} C.~P.,  {Dominik} C.,   {Natta} A.,  2001, \mn@doi [\apj]
  {10.1086/323057}, \href
  {https://ui.adsabs.harvard.edu/\#abs/2001ApJ...560..957D} {560, 957}

\bibitem[\protect\citeauthoryear{{Dullemond} et~al.,}{{Dullemond}
  et~al.}{2018}]{Dullemond2018}
{Dullemond} C.~P.,  et~al., 2018, \mn@doi [\apjl] {10.3847/2041-8213/aaf742},
  \href {https://ui.adsabs.harvard.edu/abs/2018ApJ...869L..46D} {869, L46}

\bibitem[\protect\citeauthoryear{{Ercolano} \& {Pascucci}}{{Ercolano} \&
  {Pascucci}}{2017}]{2017RSOS....470114E}
{Ercolano} B.,  {Pascucci} I.,  2017, \mn@doi [Royal Society Open Science]
  {10.1098/rsos.170114}, \href
  {https://ui.adsabs.harvard.edu/abs/2017RSOS....470114E} {4, 170114}

\bibitem[\protect\citeauthoryear{{Espaillat} et~al.,}{{Espaillat}
  et~al.}{2014}]{Espaillat2014}
{Espaillat} C.,  et~al., 2014, in {Beuther} H.,  {Klessen} R.~S.,  {Dullemond}
  C.~P.,   {Henning} T.,  eds, Protostars and Planets VI. p.~497 (\mn@eprint
  {arXiv} {1402.7103}), \mn@doi{10.2458/azu_uapress_9780816531240-ch022}

\bibitem[\protect\citeauthoryear{{Facchini}, {Pinilla}, {van Dishoeck}  \& {de
  Juan Ovelar}}{{Facchini} et~al.}{2018}]{Facchini2018}
{Facchini} S.,  {Pinilla} P.,  {van Dishoeck} E.~F.,   {de Juan Ovelar} M.,
  2018, \mn@doi [\aap] {10.1051/0004-6361/201731390}, \href
  {https://ui.adsabs.harvard.edu/abs/2018A&A...612A.104F} {612, A104}

\bibitem[\protect\citeauthoryear{{Fairlamb}, {Oudmaijer}, {Mendigut{\'\i}a},
  {Ilee}  \& {van den Ancker}}{{Fairlamb} et~al.}{2015}]{Fairlamb}
{Fairlamb} J.~R.,  {Oudmaijer} R.~D.,  {Mendigut{\'\i}a} I.,  {Ilee} J.~D.,
  {van den Ancker} M.~E.,  2015, \mn@doi [\mnras] {10.1093/mnras/stv1576},
  \href {https://ui.adsabs.harvard.edu/abs/2015MNRAS.453..976F} {453, 976}

\bibitem[\protect\citeauthoryear{{Flock}, {Fromang}, {Gonz{\'a}lez}  \&
  {Commer{\c{c}}on}}{{Flock} et~al.}{2013}]{Flock2013}
{Flock} M.,  {Fromang} S.,  {Gonz{\'a}lez} M.,   {Commer{\c{c}}on} B.,  2013,
  \mn@doi [\aap] {10.1051/0004-6361/201322451}, \href
  {https://ui.adsabs.harvard.edu/abs/2013A&A...560A..43F} {560, A43}

\bibitem[\protect\citeauthoryear{{Forgan}, {Ilee}  \& {Meru}}{{Forgan}
  et~al.}{2018}]{2018ApJ...860L...5F}
{Forgan} D.~H.,  {Ilee} J.~D.,   {Meru} F.,  2018, \mn@doi [\apj]
  {10.3847/2041-8213/aac7c9}, \href
  {https://ui.adsabs.harvard.edu/abs/2018ApJ...860L...5F} {860, L5}

\bibitem[\protect\citeauthoryear{{Gaia Collaboration} et~al.,}{{Gaia
  Collaboration} et~al.}{2016}]{2016A&A...595A...1G}
{Gaia Collaboration} et~al., 2016, \mn@doi [\aap]
  {10.1051/0004-6361/201629272}, \href
  {https://ui.adsabs.harvard.edu/abs/2016A%26A...595A...1G} {595, A1}

\bibitem[\protect\citeauthoryear{{Gaia Collaboration} et~al.,}{{Gaia
  Collaboration} et~al.}{2018}]{2018A&A...616A...1G}
{Gaia Collaboration} et~al., 2018, \mn@doi [\aap]
  {10.1051/0004-6361/201833051}, \href
  {https://ui.adsabs.harvard.edu/abs/2018A%26A...616A...1G} {616, A1}

\bibitem[\protect\citeauthoryear{{Garufi} et~al.,}{{Garufi}
  et~al.}{2013}]{garufi_2013}
{Garufi} A.,  et~al., 2013, \mn@doi [A\&A] {10.1051/0004-6361/201322429}, \href
  {http://adsabs.harvard.edu/abs/2013A%26A...560A.105G} {560, A105}

\bibitem[\protect\citeauthoryear{{Goldreich} \& {Tremaine}}{{Goldreich} \&
  {Tremaine}}{1979}]{goldreich_1979}
{Goldreich} P.,  {Tremaine} S.,  1979, \mn@doi [\apj] {10.1086/157448}, \href
  {http://adsabs.harvard.edu/abs/1979ApJ...233..857G} {233, 857}

\bibitem[\protect\citeauthoryear{{Grady} et~al.,}{{Grady}
  et~al.}{2013}]{2013ApJ...762...48G}
{Grady} C.~A.,  et~al., 2013, \mn@doi [\apj] {10.1088/0004-637X/762/1/48},
  \href {https://ui.adsabs.harvard.edu/abs/2013ApJ...762...48G} {762, 48}

\bibitem[\protect\citeauthoryear{{Hall}, {Forgan}, {Rice}, {Harries},
  {Klaassen}  \& {Biller}}{{Hall} et~al.}{2016}]{Hall2016}
{Hall} C.,  {Forgan} D.,  {Rice} K.,  {Harries} T.~J.,  {Klaassen} P.~D.,
  {Biller} B.,  2016, \mn@doi [\mnras] {10.1093/mnras/stw296}, \href
  {https://ui.adsabs.harvard.edu/#abs/2016MNRAS.458..306H} {458, 306}

\bibitem[\protect\citeauthoryear{{Hall}, {Rice}, {Dipierro}, {Forgan},
  {Harries}  \& {Alexander}}{{Hall} et~al.}{2018}]{2018MNRAS.477.1004H}
{Hall} C.,  {Rice} K.,  {Dipierro} G.,  {Forgan} D.,  {Harries} T.,
  {Alexander} R.,  2018, \mn@doi [\mnras] {10.1093/mnras/sty550}, \href
  {https://ui.adsabs.harvard.edu/#abs/2018MNRAS.477.1004H} {477, 1004}

\bibitem[\protect\citeauthoryear{{Huang} et~al.,}{{Huang}
  et~al.}{2018a}]{Huang2018TwHya}
{Huang} J.,  et~al., 2018a, \mn@doi [\apj] {10.3847/1538-4357/aaa1e7}, \href
  {https://ui.adsabs.harvard.edu/abs/2018ApJ...852..122H} {852, 122}

\bibitem[\protect\citeauthoryear{{Huang} et~al.,}{{Huang}
  et~al.}{2018b}]{2018ApJ...869L..42H}
{Huang} J.,  et~al., 2018b, \mn@doi [\apj] {10.3847/2041-8213/aaf740}, \href
  {https://ui.adsabs.harvard.edu/\#abs/2018ApJ...869L..42H} {869, L42}

\bibitem[\protect\citeauthoryear{{Huang} et~al.,}{{Huang}
  et~al.}{2018c}]{huang2018b}
{Huang} J.,  et~al., 2018c, \mn@doi [\apj] {10.3847/2041-8213/aaf7a0}, \href
  {https://ui.adsabs.harvard.edu/abs/2018ApJ...869L..43H} {869, L43}

\bibitem[\protect\citeauthoryear{{Juh{\'a}sz} \& {Rosotti}}{{Juh{\'a}sz} \&
  {Rosotti}}{2018}]{juhasz_2018}
{Juh{\'a}sz} A.,  {Rosotti} G.~P.,  2018, \mn@doi [\mnras]
  {10.1093/mnrasl/slx182}, \href
  {http://adsabs.harvard.edu/abs/2018MNRAS.474L..32J} {474, L32}

\bibitem[\protect\citeauthoryear{{Juh{\'a}sz}, {Benisty}, {Pohl}, {Dullemond},
  {Dominik}  \& {Paardekooper}}{{Juh{\'a}sz} et~al.}{2015}]{juhasz_2015}
{Juh{\'a}sz} A.,  {Benisty} M.,  {Pohl} A.,  {Dullemond} C.~P.,  {Dominik} C.,
   {Paardekooper} S.-J.,  2015, \mn@doi [\mnras] {10.1093/mnras/stv1045}, \href
  {http://adsabs.harvard.edu/abs/2015MNRAS.451.1147J} {451, 1147}

\bibitem[\protect\citeauthoryear{{Kama}, {Pinilla}  \& {Heays}}{{Kama}
  et~al.}{2016}]{2016A&A...593L..20K}
{Kama} M.,  {Pinilla} P.,   {Heays} A.~N.,  2016, \mn@doi [\aap]
  {10.1051/0004-6361/201628924}, \href
  {https://ui.adsabs.harvard.edu/#abs/2016A&A...593L..20K} {593, L20}

\bibitem[\protect\citeauthoryear{{Keppler} et~al.,}{{Keppler}
  et~al.}{2019}]{2019arXiv190207639K}
{Keppler} M.,  et~al., 2019, arXiv e-prints, \href
  {https://ui.adsabs.harvard.edu/abs/2019arXiv190207639K} {p. arXiv:1902.07639}

\bibitem[\protect\citeauthoryear{{Kraus} et~al.,}{{Kraus}
  et~al.}{2017}]{Kraus2017}
{Kraus} S.,  et~al., 2017, \mn@doi [\apj] {10.3847/2041-8213/aa8edc}, \href
  {https://ui.adsabs.harvard.edu/abs/2017ApJ...848L..11K} {848, L11}

\bibitem[\protect\citeauthoryear{{Kurtovic} et~al.,}{{Kurtovic}
  et~al.}{2018}]{2018ApJ...869L..44K}
{Kurtovic} N.~T.,  et~al., 2018, \mn@doi [\apj] {10.3847/2041-8213/aaf746},
  \href {https://ui.adsabs.harvard.edu/abs/2018ApJ...869L..44K} {869, L44}

\bibitem[\protect\citeauthoryear{{Lambrechts}, {Johansen}  \&
  {Morbidelli}}{{Lambrechts} et~al.}{2014}]{Lambrechts2014}
{Lambrechts} M.,  {Johansen} A.,   {Morbidelli} A.,  2014, \mn@doi [\aap]
  {10.1051/0004-6361/201423814}, \href
  {https://ui.adsabs.harvard.edu/abs/2014A&A...572A..35L} {572, A35}

\bibitem[\protect\citeauthoryear{{Lee} \& {Gu}}{{Lee} \& {Gu}}{2015}]{Lee2015}
{Lee} W.-K.,  {Gu} P.-G.,  2015, \mn@doi [\apj] {10.1088/0004-637X/814/1/72},
  \href {https://ui.adsabs.harvard.edu/#abs/2015ApJ...814...72L} {814, 72}

\bibitem[\protect\citeauthoryear{{Long} et~al.,}{{Long}
  et~al.}{2017}]{2017ApJ...838...62L}
{Long} Z.~C.,  et~al., 2017, \mn@doi [\apj] {10.3847/1538-4357/aa64da}, \href
  {https://ui.adsabs.harvard.edu/#abs/2017ApJ...838...62L} {838, 62}

\bibitem[\protect\citeauthoryear{{Long} et~al.,}{{Long}
  et~al.}{2018}]{2018ApJ...869...17L}
{Long} F.,  et~al., 2018, \mn@doi [\apj] {10.3847/1538-4357/aae8e1}, \href
  {https://ui.adsabs.harvard.edu/\#abs/2018ApJ...869...17L} {869, 17}

\bibitem[\protect\citeauthoryear{{Mathis}, {Rumpl}  \& {Nordsieck}}{{Mathis}
  et~al.}{1977}]{1977ApJ...217..425M}
{Mathis} J.~S.,  {Rumpl} W.,   {Nordsieck} K.~H.,  1977, \mn@doi [\apj]
  {10.1086/155591}, \href
  {https://ui.adsabs.harvard.edu/abs/1977ApJ...217..425M} {217, 425}

\bibitem[\protect\citeauthoryear{{Maud} et~al.,}{{Maud}
  et~al.}{2019}]{Maud2019}
{Maud} L.~T.,  et~al., 2019, arXiv e-prints, \href
  {https://ui.adsabs.harvard.edu/abs/2019arXiv190606548M} {p. arXiv:1906.06548}

\bibitem[\protect\citeauthoryear{{McMullin}, {Waters}, {Schiebel}, {Young}  \&
  {Golap}}{{McMullin} et~al.}{2007}]{2007ASPC..376..127M}
{McMullin} J.~P.,  {Waters} B.,  {Schiebel} D.,  {Young} W.,   {Golap} K.,
  2007, in {Shaw} R.~A.,  {Hill} F.,   {Bell} D.~J.,  eds,  Astronomical
  Society of the Pacific Conference Series Vol. 376, Astronomical Data Analysis
  Software and Systems XVI. p.~127

\bibitem[\protect\citeauthoryear{{Meru}, {Juh{\'a}sz}, {Ilee}, {Clarke},
  {Rosotti}  \& {Booth}}{{Meru} et~al.}{2017}]{meru_2017}
{Meru} F.,  {Juh{\'a}sz} A.,  {Ilee} J.~D.,  {Clarke} C.~J.,  {Rosotti} G.~P.,
   {Booth} R.~A.,  2017, \mn@doi [\apjl] {10.3847/2041-8213/aa6837}, \href
  {http://adsabs.harvard.edu/abs/2017ApJ...839L..24M} {839, L24}

\bibitem[\protect\citeauthoryear{{Min}, {Stolker}, {Dominik}  \&
  {Benisty}}{{Min} et~al.}{2017}]{2017A&A...604L..10M}
{Min} M.,  {Stolker} T.,  {Dominik} C.,   {Benisty} M.,  2017, \mn@doi [\aap]
  {10.1051/0004-6361/201730949}, \href
  {https://ui.adsabs.harvard.edu/abs/2017A&A...604L..10M} {604, L10}

\bibitem[\protect\citeauthoryear{{Montesinos}, {Perez}, {Casassus}, {Marino},
  {Cuadra}  \& {Christiaens}}{{Montesinos} et~al.}{2016}]{Montesinos2016}
{Montesinos} M.,  {Perez} S.,  {Casassus} S.,  {Marino} S.,  {Cuadra} J.,
  {Christiaens} V.,  2016, \mn@doi [\apj] {10.3847/2041-8205/823/1/L8}, \href
  {https://ui.adsabs.harvard.edu/abs/2016ApJ...823L...8M} {823, L8}

\bibitem[\protect\citeauthoryear{{Ogilvie} \& {Lubow}}{{Ogilvie} \&
  {Lubow}}{2002}]{ogilvie_2002}
{Ogilvie} G.~I.,  {Lubow} S.~H.,  2002, \mn@doi [MNRAS]
  {10.1046/j.1365-8711.2002.05148.x}, \href
  {http://adsabs.harvard.edu/abs/2002MNRAS.330..950O} {330, 950}

\bibitem[\protect\citeauthoryear{{Owen} \& {Lai}}{{Owen} \&
  {Lai}}{2017}]{2017MNRAS.469.2834O}
{Owen} J.~E.,  {Lai} D.,  2017, \mn@doi [\mnras] {10.1093/mnras/stx1033}, \href
  {https://ui.adsabs.harvard.edu/abs/2017MNRAS.469.2834O} {469, 2834}

\bibitem[\protect\citeauthoryear{{Owen}, {Ercolano}  \& {Clarke}}{{Owen}
  et~al.}{2011}]{Owen2011}
{Owen} J.~E.,  {Ercolano} B.,   {Clarke} C.~J.,  2011, \mn@doi [\mnras]
  {10.1111/j.1365-2966.2010.17818.x}, \href
  {https://ui.adsabs.harvard.edu/abs/2011MNRAS.412...13O} {412, 13}

\bibitem[\protect\citeauthoryear{{P{\'e}rez} et~al.,}{{P{\'e}rez}
  et~al.}{2016}]{perez_2016}
{P{\'e}rez} L.~M.,  et~al., 2016, \mn@doi [Science] {10.1126/science.aaf8296},
  \href {http://adsabs.harvard.edu/abs/2016Sci...353.1519P} {353, 1519}

\bibitem[\protect\citeauthoryear{{Pinilla}, {Benisty}  \&
  {Birnstiel}}{{Pinilla} et~al.}{2012}]{Pinilla2012}
{Pinilla} P.,  {Benisty} M.,   {Birnstiel} T.,  2012, \mn@doi [\aap]
  {10.1051/0004-6361/201219315}, \href
  {https://ui.adsabs.harvard.edu/abs/2012A&A...545A..81P} {545, A81}

\bibitem[\protect\citeauthoryear{{Pohl}, {Pinilla}, {Benisty}, {Ataiee},
  {Juh{\'a}sz}, {Dullemond}, {Van Boekel}  \& {Henning}}{{Pohl}
  et~al.}{2015}]{pohl_2015}
{Pohl} A.,  {Pinilla} P.,  {Benisty} M.,  {Ataiee} S.,  {Juh{\'a}sz} A.,
  {Dullemond} C.~P.,  {Van Boekel} R.,   {Henning} T.,  2015, \mn@doi [\mnras]
  {10.1093/mnras/stv1746}, \href
  {http://adsabs.harvard.edu/abs/2015MNRAS.453.1768P} {453, 1768}

\bibitem[\protect\citeauthoryear{{Rafikov}}{{Rafikov}}{2002}]{rafikov_2002a}
{Rafikov} R.~R.,  2002, \mn@doi [ApJ] {10.1086/339399}, \href
  {http://adsabs.harvard.edu/abs/2002ApJ...569..997R} {569, 997}

\bibitem[\protect\citeauthoryear{{Rice}, {Armitage}, {Bate}  \&
  {Bonnell}}{{Rice} et~al.}{2003}]{rice_2003}
{Rice} W.~K.~M.,  {Armitage} P.~J.,  {Bate} M.~R.,   {Bonnell} I.~A.,  2003,
  \mn@doi [\mnras] {10.1046/j.1365-8711.2003.06253.x}, \href
  {https://ui.adsabs.harvard.edu/#abs/2003MNRAS.339.1025R} {339, 1025}

\bibitem[\protect\citeauthoryear{{Rice}, {Armitage}, {Wood}  \&
  {Lodato}}{{Rice} et~al.}{2006}]{Rice2006}
{Rice} W.~K.~M.,  {Armitage} P.~J.,  {Wood} K.,   {Lodato} G.,  2006, \mn@doi
  [\mnras] {10.1111/j.1365-2966.2006.11113.x}, \href
  {https://ui.adsabs.harvard.edu/abs/2006MNRAS.373.1619R} {373, 1619}

\bibitem[\protect\citeauthoryear{{Rosenfeld}, {Andrews}, {Hughes}, {Wilner}  \&
  {Qi}}{{Rosenfeld} et~al.}{2013}]{rosenfeld_2013}
{Rosenfeld} K.~A.,  {Andrews} S.~M.,  {Hughes} A.~M.,  {Wilner} D.~J.,   {Qi}
  C.,  2013, \mn@doi [\apj] {10.1088/0004-637X/774/1/16}, \href
  {http://adsabs.harvard.edu/abs/2013ApJ...774...16R} {774, 16}

\bibitem[\protect\citeauthoryear{{Rosotti} \& {Clarke}}{{Rosotti} \&
  {Clarke}}{2018}]{RosottiClarke2018}
{Rosotti} G.~P.,  {Clarke} C.~J.,  2018, \mn@doi [\mnras]
  {10.1093/mnras/stx2769}, \href
  {https://ui.adsabs.harvard.edu/abs/2018MNRAS.473.5630R} {473, 5630}

\bibitem[\protect\citeauthoryear{{Rosotti}, {Juhasz}, {Booth}  \&
  {Clarke}}{{Rosotti} et~al.}{2016}]{Rosotti2016}
{Rosotti} G.~P.,  {Juhasz} A.,  {Booth} R.~A.,   {Clarke} C.~J.,  2016, \mn@doi
  [\mnras] {10.1093/mnras/stw691}, \href
  {https://ui.adsabs.harvard.edu/abs/2016MNRAS.459.2790R} {459, 2790}

\bibitem[\protect\citeauthoryear{{Shakura} \& {Sunyaev}}{{Shakura} \&
  {Sunyaev}}{1973}]{shakurasunyaev}
{Shakura} N.~I.,  {Sunyaev} R.~A.,  1973, \aap, \href
  {http://adsabs.harvard.edu/abs/1973A%26A....24..337S} {24, 337}

\bibitem[\protect\citeauthoryear{{Stolker} et~al.,}{{Stolker}
  et~al.}{2016a}]{stolker_2016}
{Stolker} T.,  et~al., 2016a, \mn@doi [\aap] {10.1051/0004-6361/201528039},
  \href {http://adsabs.harvard.edu/abs/2016A%26A...595A.113S} {595, A113}

\bibitem[\protect\citeauthoryear{{Stolker}, {Dominik}, {Min}, {Garufi},
  {Mulders}  \& {Avenhaus}}{{Stolker} et~al.}{2016b}]{2016A&A...596A..70S}
{Stolker} T.,  {Dominik} C.,  {Min} M.,  {Garufi} A.,  {Mulders} G.~D.,
  {Avenhaus} H.,  2016b, \mn@doi [\aap] {10.1051/0004-6361/201629098}, \href
  {https://ui.adsabs.harvard.edu/\#abs/2016A&A...596A..70S} {596, A70}

\bibitem[\protect\citeauthoryear{{Stone} \& {Norman}}{{Stone} \&
  {Norman}}{1992}]{1992ApJS...80..753S}
{Stone} J.~M.,  {Norman} M.~L.,  1992, \mn@doi [The Astrophysical Journal
  Supplement Series] {10.1086/191680}, \href
  {https://ui.adsabs.harvard.edu/#abs/1992ApJS...80..753S} {80, 753}

\bibitem[\protect\citeauthoryear{{Tang}, {Guilloteau}, {Pi{\'e}tu}, {Dutrey},
  {Ohashi}  \& {Ho}}{{Tang} et~al.}{2012}]{Tang2012}
{Tang} Y.~W.,  {Guilloteau} S.,  {Pi{\'e}tu} V.,  {Dutrey} A.,  {Ohashi} N.,
  {Ho} P.~T.~P.,  2012, \mn@doi [\aap] {10.1051/0004-6361/201219414}, \href
  {https://ui.adsabs.harvard.edu/abs/2012A&A...547A..84T} {547, A84}

\bibitem[\protect\citeauthoryear{{Teague} \& {Foreman-Mackey}}{{Teague} \&
  {Foreman-Mackey}}{2018}]{bettermoments}
{Teague} R.,  {Foreman-Mackey} D.,  2018, \mn@doi [Research Notes of the
  American Astronomical Society] {10.3847/2515-5172/aae265}, \href
  {https://ui.adsabs.harvard.edu/abs/2018RNAAS...2c.173T} {2, 173}

\bibitem[\protect\citeauthoryear{{Teague}, {Bae}, {Birnstiel}  \&
  {Bergin}}{{Teague} et~al.}{2018}]{2018ApJ...868..113T}
{Teague} R.,  {Bae} J.,  {Birnstiel} T.,   {Bergin} E.~A.,  2018, \mn@doi
  [\apj] {10.3847/1538-4357/aae836}, \href
  {https://ui.adsabs.harvard.edu/abs/2018ApJ...868..113T} {868, 113}

\bibitem[\protect\citeauthoryear{{Toomre}}{{Toomre}}{1964}]{toomre_1964}
{Toomre} A.,  1964, \mn@doi [\apj] {10.1086/147861}, \href
  {http://adsabs.harvard.edu/abs/1964ApJ...139.1217T} {139, 1217}

\bibitem[\protect\citeauthoryear{{Wagner}, {Apai}, {Kasper}  \&
  {Robberto}}{{Wagner} et~al.}{2015}]{wagner_2015}
{Wagner} K.,  {Apai} D.,  {Kasper} M.,   {Robberto} M.,  2015, \mn@doi [\apjl]
  {10.1088/2041-8205/813/1/L2}, \href
  {http://adsabs.harvard.edu/abs/2015ApJ...813L...2W} {813, L2}

\bibitem[\protect\citeauthoryear{{Wagner} et~al.,}{{Wagner}
  et~al.}{2018}]{2018ApJ...854..130W}
{Wagner} K.,  et~al., 2018, \mn@doi [\apj] {10.3847/1538-4357/aaa767}, \href
  {https://ui.adsabs.harvard.edu/#abs/2018ApJ...854..130W} {854, 130}

\bibitem[\protect\citeauthoryear{{Weingartner} \& {Draine}}{{Weingartner} \&
  {Draine}}{2001}]{Weingartner2001}
{Weingartner} J.~C.,  {Draine} B.~T.,  2001, \mn@doi [\apj] {10.1086/318651},
  \href {https://ui.adsabs.harvard.edu/abs/2001ApJ...548..296W} {548, 296}

\bibitem[\protect\citeauthoryear{{Zhang} et~al.,}{{Zhang}
  et~al.}{2018}]{Zhang2018}
{Zhang} S.,  et~al., 2018, \mn@doi [\apj] {10.3847/2041-8213/aaf744}, \href
  {https://ui.adsabs.harvard.edu/abs/2018ApJ...869L..47Z} {869, L47}

\bibitem[\protect\citeauthoryear{{Zhu}, {Dong}, {Stone}  \& {Rafikov}}{{Zhu}
  et~al.}{2015}]{Zhu2015}
{Zhu} Z.,  {Dong} R.,  {Stone} J.~M.,   {Rafikov} R.~R.,  2015, \mn@doi [\apj]
  {10.1088/0004-637X/813/2/88}, \href
  {https://ui.adsabs.harvard.edu/#abs/2015ApJ...813...88Z} {813, 88}

\bibitem[\protect\citeauthoryear{{van der Marel} et~al.,}{{van der Marel}
  et~al.}{2013}]{van_der_marel_2013}
{van der Marel} N.,  et~al., 2013, \mn@doi [Science] {10.1126/science.1236770},
  \href {http://adsabs.harvard.edu/abs/2013Sci...340.1199V} {340, 1199}

\bibitem[\protect\citeauthoryear{{van der Plas} et~al.,}{{van der Plas}
  et~al.}{2019}]{vanderplas2019}
{van der Plas} G.,  et~al., 2019, \mn@doi [\aap] {10.1051/0004-6361/201834134},
  \href {https://ui.adsabs.harvard.edu/abs/2019A&A...624A..33V} {624, A33}

\makeatother
\end{thebibliography}



\appendix

\section{Sub-mm counterparts of the scattered light shadows}
\label{sec:shadows}

\begin{figure}
\includegraphics[width=\columnwidth]{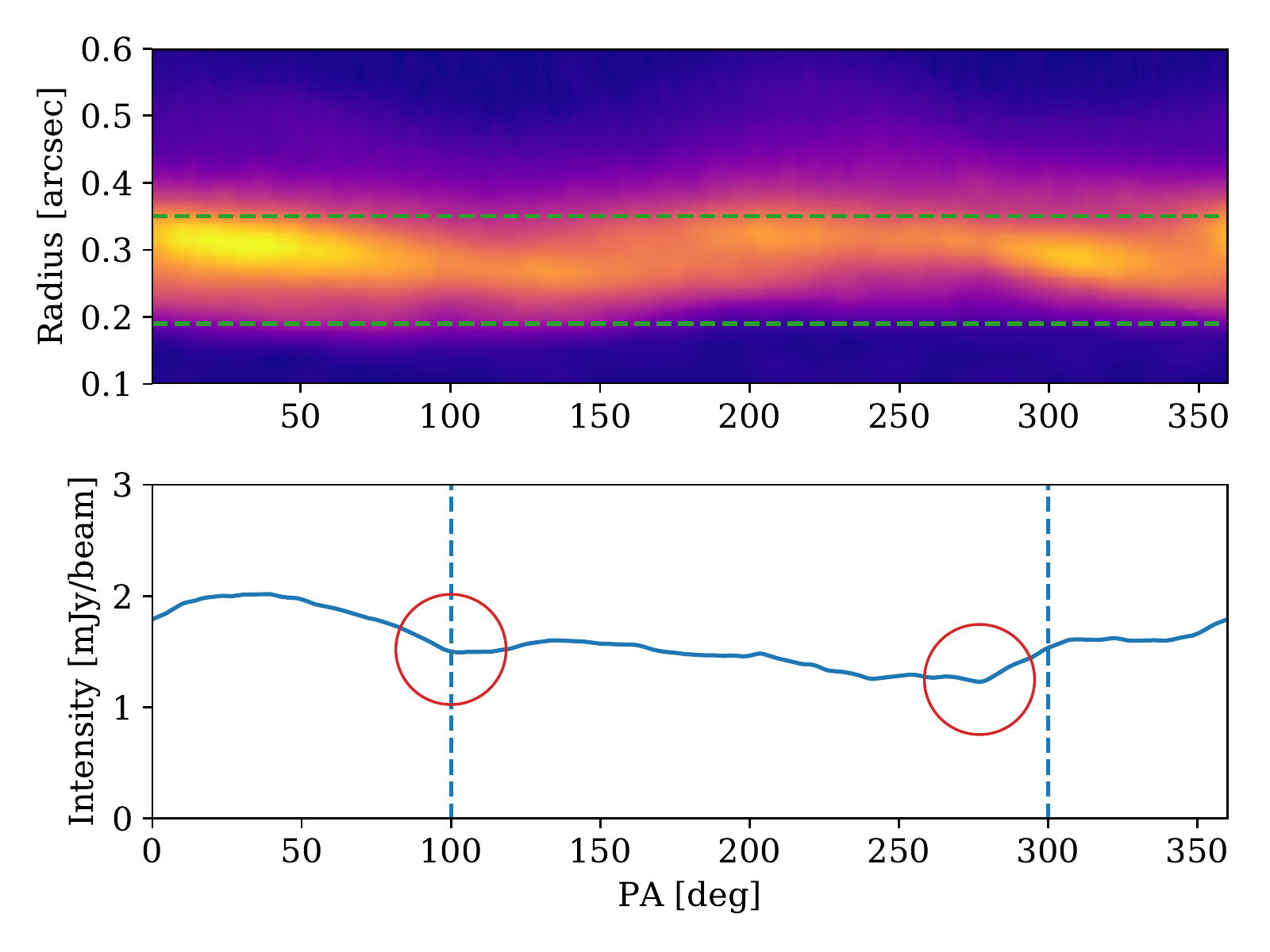}
\caption{Top panel: sub-mm continuum image after de-projection (no high-pass filtering has been applied). The dashed horizontal lines show the interval used for radial averaging. Bottom panel: azimuthal profile of the emission. The dashed vertical lines mark the position of the shadows in the scattered light image, while the circles mark the possible candidates for the sub-mm shadows counterparts. Regardless of whether these candidates are genuine, the image shows that the scattered light shadows do not correspond to significant temperature drops. We do not plot errors bars in this plot because we find that, due to large variation along radius in each azimuthal bin, the standard deviation in each bin depends very sensitively on the exact radial range used for averaging.}
\label{fig:shadows}
\end{figure}

In this section we study whether there are sub-mm counterparts of the shadows observed in the NIR scattered light, likely caused by a misaligned inner disc \citep{benisty_2017,2017A&A...604L..10M}. Figure \ref{fig:shadows} shows in the top panel the sub-mm continuum image after de-projection. In contrast to figure \ref{fig:model_comp}, here we have not made use of the high-pass filter, which enhances details on small scales and therefore would not allow us to estimate the amplitude of the shadows. The bottom panel shows the azimuthal profile of the image, averaged between the two radii indicated by the dashed green lines in the top panel.

The scattered light shadows are located at position angles 100 and 300 \degr; we mark these locations with the vertical dashed lines in the bottom panel. The image and the azimuthal profile shows a possible hint of a counter-part for the shadow at PA=100\degr, though the amplitude of this feature is quite small (less than 10 per cent). There is no such feature at PA=300, although we could tentatively identify a candidate at slightly smaller PA. We mark these two features with red circles. Regardless of whether these two features are or are not the counterparts of the scattered light shadows, their low amplitude clearly shows that the shadows do not correspond to significant temperature drops.

\bsp	
\label{lastpage}
\end{document}